\documentclass{aa}  

\usepackage{graphicx}
\usepackage{txfonts}
\usepackage{amstext}
\usepackage{xcolor}
\usepackage{threeparttable}
\usepackage{overpic}
\usepackage{multirow}
\usepackage{rotating}
\usepackage{amssymb}
\usepackage{ulem}
\usepackage{tabularx}
    \newcolumntype{L}{>{\raggedright\arraybackslash}X}
\usepackage{hyperref}
\usepackage{siunitx}
\usepackage{lscape}

\hypersetup{
  colorlinks   = true, 
  urlcolor     = blue, 
  linkcolor    = blue, 
  citecolor   = blue 
}

\begin{document}

   \title{
   Structure of the accretion flow of IX Velorum as revealed by high-resolution spectroscopy\thanks{
   Based on observations obtained at the European Southern Observatory.
   }}

   \author{J. K\'{a}ra \inst{1}
           \and L. Schmidtobreick \inst{2} 
           \and A. F. Pala \inst{3}
           \and C. Tappert \inst{4} 
            }

   \institute{Astronomical Institute, Faculty of Mathematics and Physics, 
           Charles University, V~Hole\v{s}ovi\v{c}k\'ach~2, CZ-180~00~Praha~8, \\
           Czech Republic
           \and European Southern Observatory, Casilla 19001, 7550000 Santiago 19, Chile
           \and European Space Agency, European Space Astronomy Centre, Camino Bajo del Castillo s/n, 28692 Villanueva de la Ca\~nada, Madrid, Spain
           \and Instituto de F\'{i}sica y Astronom\'{i}a de la Universidad de Valpara\'{i}so, Av. Gran Breta\~{n}a 1111, Valpara\'{i}so, Chile
           }

   \date{Version: \today}

 
  \abstract
   {Several high-mass transfer cataclysmic variables show evidence for outflow from the system, which could play an important role in their evolution. We investigate the system IX~Vel, which was proposed to show similar characteristics. }
   {We study the structure of the IX~Vel system, particularly the structure of its accretion flow and accretion disc.}
   {We use high-resolution time-resolved spectroscopy to construct radial velocity curves of the components in  IX~Vel, we compute Doppler maps of the system which we use to estimate the temperature distribution maps.}
   {
   We improve the spectroscopic ephemeris of the system and its orbital period $P_{\mathrm{orb}}=0.19392793(3)\,\mathrm{d}$. We construct Doppler maps of the system based on hydrogen and helium emission lines and the Bowen blend. The maps show features corresponding to the irradiated face of the secondary star, the outer rim of the accretion disc, and low-velocity components located outside the accretion disc and reaching towards $\mathrm{L}_3$. We constructed a temperature distribution map of the system using the Doppler maps of Balmer lines. Apart from the features found in the Doppler maps, the temperature distribution map shows a region of high temperature in the accretion disc connecting the expected position of a bright spot and the inner parts of the disc. 
   }
   {
    We interpret the low-velocity emission found in the Doppler map as emission originating in the accretion disc wind and in an outflow region located in the vicinity of the third Lagrangian point $\mathrm{L}_3$. This makes IX Vel a member of the RW Sex class of Cataclysmic Variables.
   }


   \maketitle
%

\section{Introduction}

Cataclysmic variables (CVs) are semidetached binaries consisting of a white dwarf (WD) primary star and a red dwarf secondary star. The red dwarf fills its Roche lobe and loses mass through the first Lagrangian point $\mathrm{L}_1$. This mass is then accreted onto the WD and forms an accretion disc when no strong magnetic field is present. A comprehensive review of CVs was published by \cite{1995cvs..book.....W}.

The accretion disc of CVs can experience thermal instabilities that lead to a transition from low to high-temperature states, which result in events of increased brightness called dwarf nova outbursts, with an amplitude of several magnitudes. In nova-like variables (NLs) the mass transfer rate is high enough for the disc to be constantly in a stable, high-temperature state and no outbursts are observed in these CVs. 

Systems with high mass transfer rates can exhibit  stellar winds originating in the inner parts of the accretion disc \citep{1982ApJ...260..716C, 2015MNRAS.450.3331M}. The winds can be inferred from the P Cygni profiles of resonance lines observed in ultraviolet spectra such as \ion{N}{v}, \ion{Si}{iv} or \ion{C}{iv} \citep{1990MNRAS.245..323W} and some systems show emission lines originating in wind regions also in optical spectra \citep{1986ApJ...302..388H, 1990ApJ...349..593M}.
Some NLs show single-peaked Balmer lines, which can be partly formed in an outflow region of the accretion disc, which lies in the plane of the disc close to the Lagrangian point $\mathrm{L}_3$ \citep{2017MNRAS.470.1960H,2020MNRAS.497.1475S}.

IX~Vel (also known as CPD $-48^{\circ}1577$) is a NL which was discovered by \cite{1982IAUC.3730....2G, 1984ApJ...276L..13G} using optical spectroscopic observations. 
It is one of the brightest CVs
($V \simeq 10 \, \mathrm{mag}$) and was therefore the subject of numerous studies. The system shows light curve variations on different time scales. \cite{1984Ap&SS..99..145W} observed long-term variations with an amplitude of $1\, \mathrm{mag}$ during a 9-year long observational run, which showed no systematic trend. Shorter variations with an amplitude of $0.2\,  \mathrm{mag}$ on a time scale of 1 hour were observed by \cite{1983MNRAS.204P..35W}, and
\cite{1984MNRAS.211..629W} observed $0.1 \, \mathrm{mag}$ flickering on time scale of minutes. Brightness variations on even shorter timescales were reported by \cite{1985MNRAS.212P...9W}, who observed $0.001 \, \mathrm{mag}$ oscillations with a period $\sim 25\, \mathrm{s}$. 
\cite{1988MNRAS.235.1385H} obtained continuous infrared (IR) photometry in $HJK$ bands, which revealed variations correlated with the orbital period, presenting two minima of different depths separated by roughly 0.5 orbital phases. \cite{1988MNRAS.235.1385H} interpreted this behaviour by ellipsoidal modulations, which produces sinusoidal brightness changes, and irradiation of the secondary component, which causes different depth of the minima with the deeper minimum occurring at orbital phase $0.0$, when the irradiated part of secondary is hidden out-of-view. The same light curves were later modelled by \cite{2007ApJ...662.1204L}, who interpreted the minimum at orbital phase $0.0$ as a partial eclipse of the rim of the accretion disc by the secondary.

\cite{2021arXiv211115145K} studied the long-term variability of IX~Vel using All-Sky Automated Survey \citep[ASAS-3,][]{1997AcA....47..467P} and All-Sky Automated Survey for Supernovae \citep[ASAS-SN,][]{2014ApJ...788...48S, 2017PASP..129j4502K} observations and found that the data show small-amplitude variations similar to dwarf novae outbursts and also standstills, suggesting that IX~Vel is a low-amplitude Z~Cam star, i.e. a dwarf nova alternating between states of outbursting activity and states of high constant brightness (standstills) similar to those in NLs.

The spectrum of IX~Vel  contains broad hydrogen absorption lines with an emission component in their core, as was already reported by \cite{1982IAUC.3730....2G} upon its discovery. Subsequent optical phase-resolved spectroscopic observations by \cite{1983MNRAS.204P..35W} allowed for the construction of the radial velocity curve and for the estimation of the orbital period $P = 0.187 \, \mathrm{d}$. 
\cite{1987MitAG..70..369B, 1990A&A...230..326B} 
studied the profiles of Balmer and \ion{He}{i} emission lines, which consist of a broad component originating in the accretion disc and a narrow component originating at the irradiated surface of the secondary.
They used the narrow emission component to measure the radial velocities of the irradiated surface of the secondary and the broad component to measure radial velocities of the inner part of the disc 
to model the irradiation-produced lines, and to determine the inclination of the system ($i = 60^{\circ} \pm 5^{\circ}$). They used the derived inclination and radial velocities to determine the masses of both components ($M_1 = 0.80^{+0.16}_{-0.11} \;\mathrm{M}_{\sun}$ and $M_2 = 0.52^{+0.10}_{-0.07} \; \mathrm{M}_{\sun}$). 

\cite{1999AcA....49...73K} obtained phase-resolved spectroscopic observations in the spectral range $8100-9200 \,$\AA{}, which showed \ion{Ca}{ii} and Pashen hydrogen emission lines. They modelled the emission lines with a three-component model consisting of a disc, a secondary, and a hot spot, whose properties differ from those of a typical bright spot in an accretion disc. The model showed that a hot spot at the rim of the accretion disc contributes to the spectrum of the system and based on the radial velocities of the hot spot, its transit occurs at the phase $\varphi = 0.19$\footnote{\cite{1999AcA....49...73K} gives a value for the transit $\varphi = 0.32$, which is however not corrected for the phase offset of the spectra, which can be determined from the phase offset of the radial velocity curves of the secondary and the disc. Here we present the corrected value. In both cases, however, the position of the hot spot component does not match the position of a typical CV bright spot.}, which means that the hot spot component does not coincide with the position where a stream of transferring mass meets the accretion disc, which is typically referred to as the hot spot in CVs.

\cite{1985ApJ...292..601S} used ultraviolet spectra obtained with the \textit{International Ultraviolet Explorer} (\textit{IUE}) to model the system and estimated a wind mass-loss rate $\dot{M}_{\mathrm{wind}} \leq 10^{-9} $ M$_{\sun}$ year$^{-1}$ and a mass accretion rate $\dot{M}_{\mathrm{acc}} \approx 5 \cdot 10^{-9} $ M$_{\sun}$ year$^{-1}$. 
The \textit{IUE} spectra were analysed also by \cite{1991ApJ...373..624M}, who determined that the velocity of the wind is lowest on the side facing towards the secondary and largest on the side facing away from the secondary.

Even though IX Vel was the subject of numerous studies in the past, none of them employed the Doppler tomography method to analyse the system. Here, we present a study of IX Vel which makes use of new time-resolved spectroscopic observations with high temporal and wavelength resolution, which we use to determine the radial velocities of various components of the system and to construct Doppler maps based on observed emission lines, which we use to describe the structure of emitting regions in the system. By combining Doppler maps of hydrogen lines, we compute a rough approximation of the temperature distribution in the system.

\section{Observations}
\label{sec:observations}

We have used spectroscopic observations obtained with different instruments and different facilities. Below, we give short descriptions of the used instruments and obtained data. A log of the observations is reported in Table~\ref{T:01}.

\subsection{HARPS}

Time-resolved spectroscopic observations were obtained with the High Accuracy Radial velocity Planet Searcher
\citep[HARPS, ][]{2003Msngr.114...20M} located at the 3.6–meter telescope on La Silla Observatory, Chile. 
HARPS is a fibre-fed, cross-dispersed {\'e}chelle spectrograph with a spectral coverage $\lambda \simeq 3800-6900\,$\AA{}. We used short exposures ($t_{\mathrm{exp}} = 300\; \mathrm{s}$ ) and the high-efficiency EGGs mode, in order to achieve a  good phase resolution and maximise the signal-to-noise ratio.
The resulting resolving power is about $R = 80\,000$.

 The observations took place during three nights, a set of 55 spectra was obtained in 2017 on January 11 and two sets of 53 and 51 spectra were obtained in 2019 on February 20 and 21, respectively. 
Each set covers a time interval of approximately one orbital period of the system $P_\mathrm{orb}=4.6$ hours. 
The data have been reduced using the standard HARPS pipeline installed on La Silla. 
All spectra have been normalised by fitting a polynomial function of the 5th order to the continuum.

\begin{table*}

    \centering
        \caption{Summary of spectroscopic observations}
    \label{T:01}
    \centering
    \begin{tabular}{l l r @{ -- } l c r @{ -- } l c c c c}
\hline \hline  \noalign{\smallskip}
         Instrument	&	Date	&	\multicolumn{2}{c}{Spectral range}	&	Slit/fibre	&	\multicolumn{2}{c}{Seeing}	&	$R =  \lambda / \Delta \lambda $	&	$\mathrm{N}_{\mathrm{exp}}$	&	$t_{\mathrm{exp}}$	&	$\Delta t$	\\
	&		&	\multicolumn{2}{c}{[\AA]}	&	arcsec	&	\multicolumn{2}{c}{arcsec}	&		&		&	[s]	&	[h]	\\
\hline		\noalign{\smallskip}															
UVES	&	2002 Jun 4	&$	3282 $ & $ 4563	$&$	0.44	$&	\multicolumn{2}{c}{1.73}	&$	49\,620	$&$	1	$&$	2700	$&$	0.75	$\\
	&		&$	4583 $ & $ 6687	$&$	0.3	$&	\multicolumn{2}{c}{1.73}	&$	51\,690	$&$	1	$&$	2700	$&$	0.75	$\\
	&	2012 Mar 5	&$	3281 $ & $ 4562	$&$	0.8	$&	\multicolumn{2}{c}{0.82}	&$	68\,642	$&$	1	$&$	120	$&$	0.03	$\\
	&		&$	4583 $ & $ 6686	$&$	0.8	$&	\multicolumn{2}{c}{0.82}	&$	107\,200	$&$	1	$&$	120	$&$	0.03	$\\
\hline			\noalign{\smallskip}														
X-shooter	&	2014 Oct 20	&$	2989 $ & $ 5560	$&$	1.3, 5.0	$&	\multicolumn{2}{c}{0.92, 0.94}	&$	2000, 1900	$&$	2	$&$	20	$&$	0.01	$\\
	&		&$	5337 $ & $ 10\,200	$&$	1.2, 5.0	$&	\multicolumn{2}{c}{0.88, 0.95}	&$	3300, 3200	$&$	2	$&$	20	$&$	0.01	$\\
	&		&$	9940 $ & $ 24\,790	$&$	1.2, 5.0	$&	\multicolumn{2}{c}{0.88, 0.95}	&$	3900, 2600	$&$	2	$&$	20	$&$	0.01	$\\
\hline			\noalign{\smallskip}														
HARPS	&	2017 Jan 11	&$	3800 $ & $ 6900	$&$	1	$&$	0.73 $ & $ 1.33	$&$	163\,000	$&$	55	$&$	300	$&$	4.97	$\\
	&	2019 Feb 20	&$	3800 $ & $ 6900	$&$	1	$&$	0.74 $ & $ 2.2	$&$	158\,000	$&$	53	$&$	300	$&$	4.87	$\\
	&	2019 Feb 21	&$	3800 $ & $ 6900	$&$	1	$&$	0.4 $ & $ 0.88	$&$	156\,000	$&$	51	$&$	300	$&$	4.60	$\\
\hline		\noalign{\smallskip}															
\textit{HST}	&	2017 Feb 11	&$	1660 $ & $ 3070	$&$	0.5	$& \multicolumn{2}{c}{} &$	910	$&$	3	$&$	105	$&$	0.15	$\\
	&		&$	2890 $ & $ 5700	$&$	0.5	$& \multicolumn{2}{c}{} &$	800	$&$	3	$&$	30	$&$	0.09	$\\
	&		&$	5260 $ & $ 10\,250	$&$	0.5	$& \multicolumn{2}{c}{} &$	790	$&$	1	$&$	60	$&$	0.05	$\\
\hline		\noalign{\smallskip}															
\textit{IUE}	&	1982 Nov 17	&$	1150 $ & $ 1950	$&$		$& \multicolumn{2}{c}{} &$	300	$&$	3	$&$	119 $ -- $ 10\,800	$&$	3.07	$\\
	&		&$	1900 $ & $ 3200	$&$		$& \multicolumn{2}{c}{} &$	300	$&$	3	$&$	70 $ -- $ 9000	$&$	2.54	$\\
	&	1983 Apr 19	&$	1150 $ & $ 1950	$&$		$& \multicolumn{2}{c}{} &$	300	$&$	6	$&$	90 $ -- $ 239	$&$	0.27	$\\
	&		&$	1900 $ & $ 3200	$&$		$& \multicolumn{2}{c}{} &$	300	$&$	4	$&$	60 $ -- $ 180	$&$	0.12	$\\
	&	1984 Feb 25	&$	1150 $ & $ 1950	$&$		$& \multicolumn{2}{c}{} &$	300	$&$	5	$&$	119 $ -- $ 3000	$&$	1.00	$\\
	&		&$	1900 $ & $ 3200	$&$		$& \multicolumn{2}{c}{} &$	300	$&$	1	$&$	180	$&$	0.05	$\\
	&	1985 Mar 14	&$	1150 $ & $ 1950	$&$		$& \multicolumn{2}{c}{} &$	300	$&$	3	$&$	2160, 3600	$&$	2.60	$\\
	&		&$	1900 $ & $ 3200	$&$		$& \multicolumn{2}{c}{} &$	300	$&$	2	$&$	1380	$&$	0.77	$\\
	&	1985 Jun 3	&$	1150 $ & $ 1950	$&$		$& \multicolumn{2}{c}{} &$	300	$&$	1	$&$	90	$&$	0.02	$\\
	&		&$	1900 $ & $ 3200	$&$		$& \multicolumn{2}{c}{} &$	300	$&$	2	$&$	90, 390	$&$	0.13	$\\
	&	1986 Nov 6-7	&$	1150 $ & $ 1950	$&$		$& \multicolumn{2}{c}{} &$	300	$&$	12	$&$	119 $ -- $ 3600	$&$	6.25	$\\
	&		&$	1900 $ & $ 3200	$&$		$& \multicolumn{2}{c}{} &$	300	$&$	7	$&$	60 $ -- $ 119	$&$	0.16	$\\
	&	1991 Nov 2-3	&$	1150 $ & $ 1950	$&$		$& \multicolumn{2}{c}{} &$	300	$&$	15	$&$	80 $ -- $ 119	$&$	0.41	$\\
    \hline
    \end{tabular}
    \tablefoot{
    The total time of observations $\Delta t$ was determined as the time elapsed between the start of the first exposure and the end of the last exposure from the corresponding run. This approach was chosen to better describe the phase coverage of the time-resolved spectroscopy. In the case of \textit{IUE} data, which were not always obtained consecutively, $\Delta t$ was determined as the sum of exposure times of all spectra in corresponding runs.
    Slit/fibre column lists the width of slits or diameters of the fibre, $\mathrm{N}_{\mathrm{exp}}$ denotes the number of exposures.
    }

\end{table*}{}

\subsection{X-shooter}

X-shooter \citep{xshooter} is an {\'e}chelle spectrograph located at the Cassegrain focus of UT3 of the Very Large Telescope (VLT) of the European Southern Observatory (ESO) on Cerro Paranal (Chile).
The instrument is equipped with three arms: blue (UVB, $\lambda \simeq 3000-5595\,$\AA), visual (VIS, $\lambda \simeq 5595-10\,240\,$\AA) and near-infrared (NIR, $\lambda \simeq 10\,240-24\,800\,$\AA), which allow to cover the wavelength range from $\simeq 3\,000\,$\AA\,up to $\simeq 25\,000\,$\AA\, in one exposure with a resolution $R \simeq 5000 - 10\,000$.
IX\,Vel was observed with X-shooter on October 20th, 2014. Spectra were obtained with slit widths of 1.3\arcsec\, in the UVB arm, 1.2\arcsec\, in the VIS arm, and 1.2\arcsec\, in the NIR arm, using an exposure time of 20\,s for each arm.
The data were reduced using the Reflex pipeline \citep{reflex} and a telluric correction was performed using molecfit \citep{molecfit1,molecfit2}.

\subsection{UVES}

We also used spectra of IX~Vel obtained with ESO's Ultraviolet and Visual Echelle Spectrograph  \citep[UVES,][]{2000SPIE.4008..534D}
and available in the ESO Science Archive Facility\footnote{Available at \url{archive.eso.org}}.
The spectra were obtained during two epochs (2002, June and 2012, March). For each run two setups were used, one covering the spectral range $3282 - 4563$ \AA{} and the other covering the spectral range $4583 - 6687$ \AA{}. The spectra were obtained with 
different exposure times in each run, see Table~\ref{T:01} for details.

\subsection{IUE}

\textit{IUE} is a satellite which was operational from 1978 to 1996. It was equipped with a $0.45$ m telescope and two {\'e}chelle spectrographs covering spectral ranges $\lambda \simeq 1150 - 1950\,$\AA{} and $\lambda \simeq 1900 - 3200\,$\AA. The spectrographs could be operated in high-resolution mode with a resolution $R \simeq 12\,000$ and in low resolution mode with a resolution $R \simeq 300$. For a full description of the \textit{IUE} satellite and its instruments, see \cite{1978Natur.275..372B} and \cite{1978Natur.275..377B}. 

IX~Vel was observed in the years 1982 - 1991 with various set-ups.
Out of 64 spectra, 15 of them were observed in high-resolution mode, the rest was observed in low resolution mode. For the purpose of this paper, we  used low-resolution spectra and high-resolution spectra which were re-sampled to a low resolution, all spectra were retrieved with ESASky\footnote{Available at \url{sky.esa.int}} \citep{2018A&C....24...97G, 2017PASP..129b8001B}.

\subsection{HST/STIS}

The Space Telescope Imaging Spectrograph (STIS) is an instrument onboard the \textit{Hubble Space Telescope} (\textit{HST}). Spectra of IX~Vel used in this paper were obtained in 2017 February 11 in three different spectral ranges. Gratings G230LB, G430L and G750L were used to obtain spectra with spectral ranges  $\lambda \simeq 1660 - 3070\,$\AA{}, $\lambda \simeq 2890 - 5700\,$\AA{} and $\lambda \simeq 5260 - 10\,250\,$\AA{}, respectively, and a spectral resolution $R \simeq 790 - 900$  The spectra were retrieved with ESASky.

\section{Spectra}

\begin{figure*}[ptbh]
        \centering
      
       \includegraphics[width=0.98\textwidth]{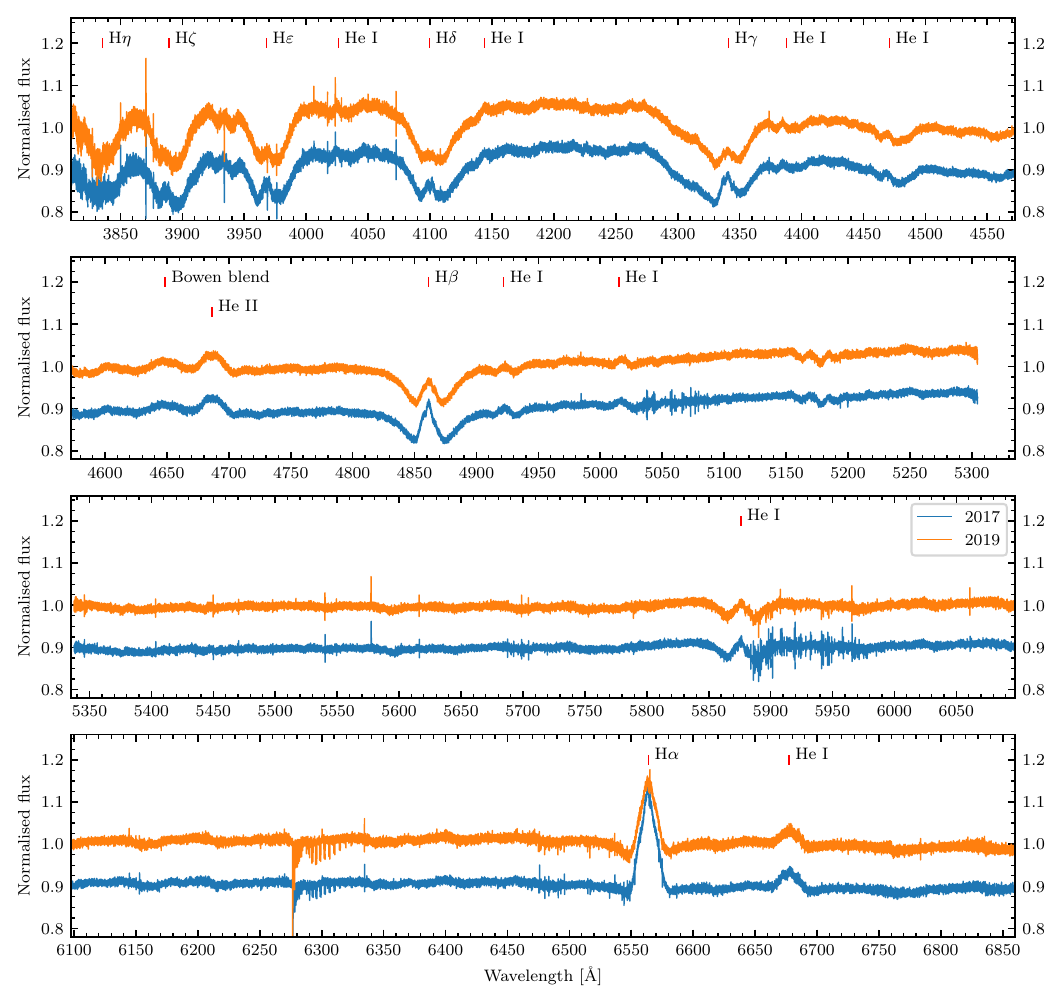}
   
        \caption{Average spectra of IX~Vel obtained with HARPS on January 11, 2017, and on February 20, 2019, the 2017 spectrum is vertically shifted for clarity. The spectrum has been corrected for telluric lines, the main emission/absorption features are labelled.}
        \label{F:SPEC_D2}
\end{figure*}

An average optical spectrum of IX~Vel constructed using the observations obtained with HARPS on February 20, 2019 is shown in Fig.~\ref{F:SPEC_D2}.
The spectrum shows broad Balmer absorption lines with multi-component emission cores. The absorption component of H$\alpha$ is shallow while for the high-order Balmer lines the emission component is becoming weaker than the absorption component and for H$\gamma$ and higher-order lines, absorption is the dominant feature. 
There are also He I $\lambda4026$, $\lambda4144$, $\lambda4388$, $\lambda4471$, $\lambda4922$, $\lambda5015$, $\lambda5876$, $\lambda6677$ emission lines accompanied by less prominent absorption component, and He II $\lambda4682$ and Bowen blend emission lines. 
While HARPS spectra obtained in 2017 and 2019 share the same general shape of features, the ones obtained in 2017 show less prominent absorption, which is most noticeable in H$\alpha$ for which the absorption component is almost absent.

\begin{figure}[tbh]
        \centering
       \includegraphics[width=0.49\textwidth]{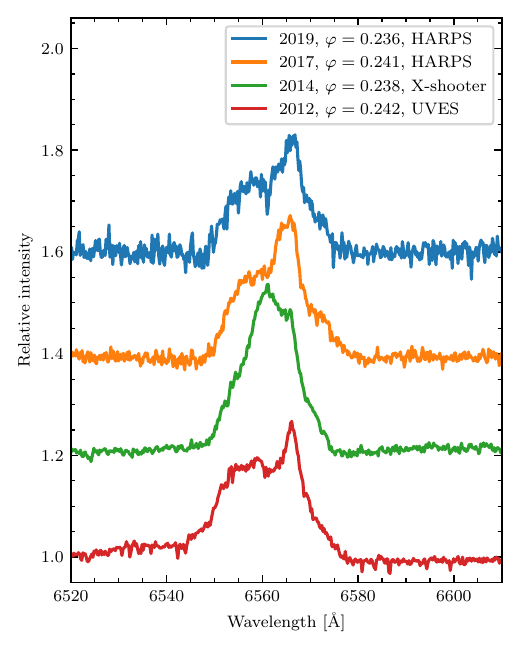}
        \caption{H$\alpha$ line observed in different years. All observations show the spectra at a similar orbital phase $\varphi \approx 0.24$. All spectra are normalised, spectra obtained with HARPS and UVES are shown binned to match the resolution of the X-shooter data. A vertical offset has been applied to the spectra for clarity.}
        \label{F:03}
\end{figure}

The Balmer emission lines consist of a broad component and a narrow component, as can be seen in Fig.~\ref{F:03}. 
The same figure also shows a comparison of the H$\alpha$ line as observed using UVES in 2012, X-shooter in 2014 and HARPS in 2017 and 2019. All observations presented in Fig.~\ref{F:03} were obtained at a similar orbital phase of $\simeq 0.24$ and the HARPS and UVES spectra were binned to the resolution of the X-shooter data, yet the shape of the line observed in 2014 differs significantly from the other three examples in that the narrow component is not the strongest feature of the emission line, instead, an emission centred on wavelength $6560\;\mathrm{ \AA}$ dominates the emission-line profile. All HARPS spectra show the same general shape of the emission lines, therefore, we have combined the 2017 and 2019 HARPS data for the analysis of the disc structure.

\begin{figure*}[tbh]
        \centering
       
       \includegraphics[width=0.98\textwidth]{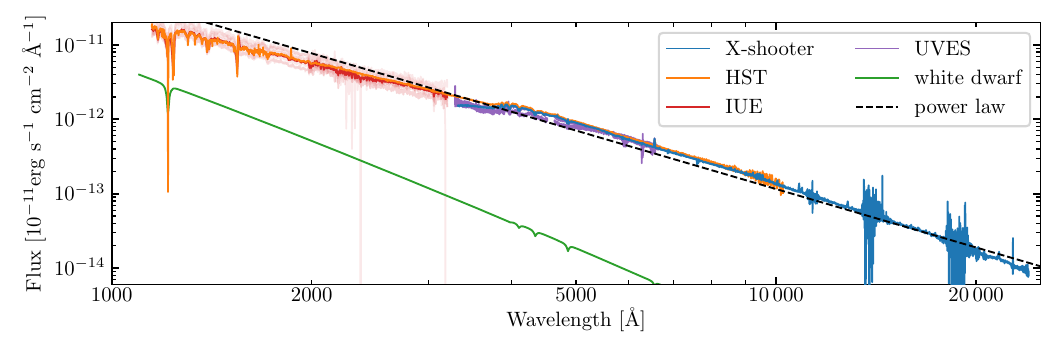}

        \caption{ SED of IX~Vel obtained with different instruments. X-shooter and \textit{HST} spectra are presented in their original flux, the flux of \textit{IUE} and UVES spectra was multiplied by an arbitrary factor so that it matches the values of the X-shooter and \textit{HST} spectra. The green line represents the expected contribution of a WD with a temperature of $T = 60\,000\;\mathrm{K}$, mass $M_{\mathrm{WD}} = 0.8 \;\mathrm{M}_{\sun}$, and radius $R_{\mathrm{WD}} = 0.015\;\mathrm{R}_{\sun}$   \citep{2007ApJ...662.1204L} and a distance of $d = 90 \; \mathrm{pc}$ \citep{2020yCat.1350....0G}. The black dashed line represent a fit to the spectrum with a power law whose slope is $-2.6$.}
        \label{F:SPEC_FULL}
\end{figure*}

The high resolution of the HARPS data allows us to study the emission lines in great detail. To inspect the general Spectral Energy Distribution (SED), one needs, however, a larger wavelength range and we combined   archival data from the various instruments mentioned in Section~\ref{sec:observations}. The result is displayed in Fig.~\ref{F:SPEC_FULL}.
 The SED 
 shows a continuous increase of the flux towards shorter wavelengths; no indication of a maximum from a Planck-curve is present. 
The spectrum follows a power law with a slope of $-2.6$. This slope is similar to the slope of a hot steady state disc, which is expected to be $-2.3$ \citep{1969Natur.223..690L}. \cite{2017ApJ...846...52G} derived slopes of IUE spectra for a set of 105 CVs including IX Vel, for which they derived value of $-2.2$, similar to other disc-dominated NLs in the sample.
 It can be therefore concluded, that the flux is dominated by the accretion disc as has been seen in other NL stars when accreting \citep[see e.g.][]{2012MNRAS.422.2332R}.
 This is similar to e.g. DW\,UMa for which the contribution of the accretion disc to the SED in the UV and the fact that the WD is completely overshone was nicely demonstrated \citep{2004ApJ...615L.129K}.

\section{Radial velocities}

The line profile consists of multiple components, each of which we studied separately. This allowed us to probe the radial velocities of the WD with different methods and to also determine the radial velocity curve of the irradiated face of the secondary. Below, we present the analysis of the radial velocities measured from the absorption components, the broad emission components and the narrow emission components observed in the HARPS data.

\subsection{Absorption features}
\label{S:AF}

\begin{figure}[!tbh]
        \centering
       \includegraphics{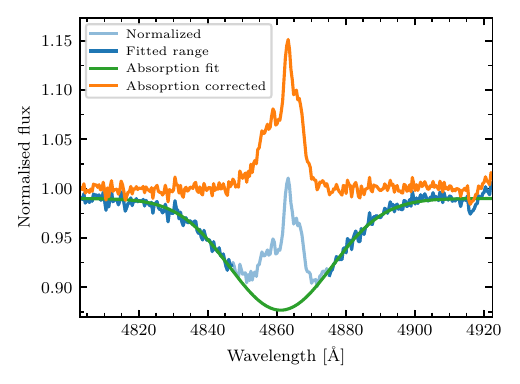}
        \caption{Sample plot of the H$\beta$ line observed on 21st February 2019 showing the fitting of the absorption line and the absorption-corrected spectra, the observational data are binned for clarity.
        }
        \label{F:HBF}
\end{figure}

To derive the radial velocities of the absorption components of H$\alpha$, H$\beta$, H$\gamma$, H$\delta$, and H$\varepsilon$ we need to minimise the influence of the emission lines within the absorption troughs. To achieve this, we opted for a procedure of fitting and cross-correlating in two steps. 
First, we constructed an average absorption line profile by fitting each line with a Gaussian and determining its average value. This average profile was then used to determine the radial velocities by cross-correlation. 
The central emission lines were masked during both steps, so the derived value is influenced only by the wings of the absorption line.
Fig.~\ref{F:HBF} shows an example of the absorption profile fitting in the case of the H$\beta$ line.
When computing the mean values, we treated spectra obtained in 2017 and 2019 separately to accommodate for the mentioned changes in the absorption profile. The cross-correlation of the absorption component of H$\alpha$ was performed only for the 2019 spectra because it was too weak in the 2017 spectra.
\begin{figure}[!tbh]
        \centering
       \includegraphics[width=0.49\textwidth]{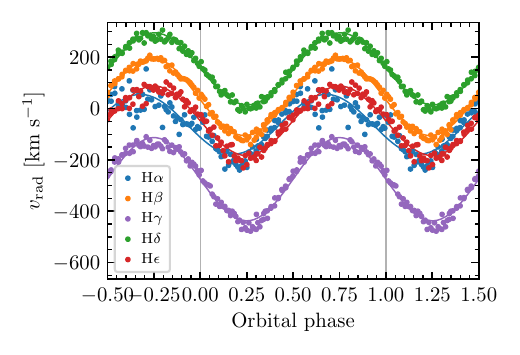}
        \caption{Radial velocity curves of the absorption components of Balmer lines observed in the HARPS spectra obtained in 2019. The solid lines show the best fitting models to the radial velocity curves, and the best fitting parameters are given in Table~\ref{T:RV_SOLO}.}
        \label{F:RV_01}
\end{figure}

\begin{table*}[ht]
        \caption{Best fitting parameters to the radial velocity curves derived from the analysis of the Balmer lines.}
    \label{T:RV_SOLO}
    \centering

        \begin{tabular}{l r r r  | r r r }
                & \multicolumn{3}{c}{2017} & \multicolumn{3}{c}{2019} \\
        \hline \noalign{\smallskip}
        Line  & \multicolumn{1}{c}{$\varphi_0$} & \multicolumn{1}{c}{$\gamma$} & \multicolumn{1}{c |}{$K$} & \multicolumn{1}{c}{$\varphi_0$} & \multicolumn{1}{c}{$\gamma$} & \multicolumn{1}{c}{$K$}  \\
         & & \multicolumn{1}{c}{[$\text{km} \cdot \text{s}^{-1}$]} & \multicolumn{1}{c |}{[$\text{km} \cdot \text{s}^{-1}$]} & & \multicolumn{1}{c}{[$\text{km} \cdot \text{s}^{-1}$]} & \multicolumn{1}{c}{[$\text{km} \cdot \text{s}^{-1}$]} \\
         \hline \hline \noalign{\smallskip}
         &\multicolumn{6}{c}{Absorption features}\\
         \hline \noalign{\smallskip}
        H$\alpha$ &  & & &                                                  $-0.072 \pm 0.008$ & $-61.3 \pm 4.0 $& $116.4 \pm 5.6$ \\
        H$\beta$ & $-0.017 \pm 0.003$ & $35.0 \pm 2.4 $& $167.7 \pm 3.3$  & $-0.012 \pm 0.002$ & $45.8 \pm 1.2 $& $151.0 \pm 1.8$ \\
        H$\gamma$ & $-0.003 \pm 0.004$ & $-504.4 \pm 3.7 $& $198.4 \pm 5.0$  & $0.000 \pm 0.003$ & $-275 \pm 2.0 $& $162.1 \pm 2.8$ \\
        H$\delta$ & $0.015 \pm 0.002$ & $163.1 \pm 1.7 $& $154.7 \pm 2.3$    & $0.003 \pm 0.002$ & $154.5 \pm 1.3 $& $140.3 \pm 1.9$ \\
        H$\varepsilon$ & $0.001 \pm 0.005$ & $-27.3 \pm 3.1 $& $140.0 \pm 4.2$  & $0.000 \pm 0.004$ & $-47.7 \pm 2.1 $& $135.6 \pm 2.9$ \\
        \hline \noalign{\smallskip}
        &\multicolumn{6}{c}{Broad emission features}\\
         \hline \noalign{\smallskip}
        H$\alpha$ & $-0.145 \pm 0.008$ & $36.3 \pm 3.1 $ & $85.4 \pm 4.4$      & $-0.041 \pm 0.003$ & $42.1 \pm 1.1 $& $77.1 \pm 1.5$ \\
        H$\beta$ & $0.085 \pm 0.010$ & $42.0 \pm 3.7 $& $84.7 \pm 5.5$        & $0.063 \pm 0.006$ & $46.4 \pm 2.6 $& $94.7 \pm 3.7$ \\
        \hline \noalign{\smallskip}
        &\multicolumn{6}{c}{Narrow emission feature}\\
        \hline \noalign{\smallskip}
         H$\alpha$\tablefootmark{a} & $0.528 \pm 0.002$ & $34.9 \pm 1.3 $ & $119.8 \pm 1.4$ &$0.528 \pm 0.002$ & $34.9 \pm 1.3 $ & $119.8 \pm 1.4$ \\

    \hline
    \end{tabular}
    
    \tablefoot{
    \tablefoottext{a}{Radial velocities from both years were fitted simultaneously in case of the narrow H$\alpha$ emission.}
    }
\end{table*}{}

We fitted each radial velocity curve with a sine function defined as follow:
\begin{equation}
    v_\mathrm{rad} = \gamma - K \sin \left(2\pi \left( \varphi - \varphi_0 \right) \right),
\end{equation}
where $v_\mathrm{rad}$ is the radial velocity, $\varphi_0$ is the zero point of the ephemeris, $\gamma$ is the systemic velocity and $K$ is the velocity amplitude. The parameters of the best-fitting models are given in Table~\ref{T:RV_SOLO}, 
Fig.~\ref{F:RV_01} shows the radial velocities obtained in 2019 phased-folded on the orbital period.

Except for H$\alpha$, all velocity curves based on the hydrogen absorption lines have a similar zero phase. H$\alpha$ is noticeably phase-shifted, as can be seen in Fig.~\ref{F:RV_01}.
Since H$\alpha$ has the weakest absorption component out of the hydrogen lines, it is likely that the velocity determination was affected by the contribution of the emission component. 

The $\gamma$-velocities  differ for the separate velocity curves between values $\sim -500 \; \mathrm{km\,s}^{-1}$ and $\sim 160 \; \mathrm{km\,s}^{-1}$. The shift of the radial velocity curve is most notable in the case of H$\gamma$, which is significantly blue-shifted with respect to the other hydrogen lines. This is most likely caused by the blending of the hydrogen absorption lines with metal lines present in the spectra, causing an asymmetric line profile around $4300\; \mathrm{\AA}$, as can be seen  in Fig.~\ref{F:SPEC_D2}.
The wavelength calibration of the spectra is robust enough for the discrepancies to be caused by calibration inaccuracies.

\subsection{Broad emission lines}
\label{S:BEL}

As an additional check,  the wings of the H$\alpha$ and H$\beta$ emission lines were used to determine radial velocities via the method of fitting a double-Gaussian profile as described by \cite{1980ApJ...238..946S} and \cite{1983ApJ...267..222S}. 

\begin{figure}[!tb]
        \centering
        
        \includegraphics[width=0.49\textwidth]{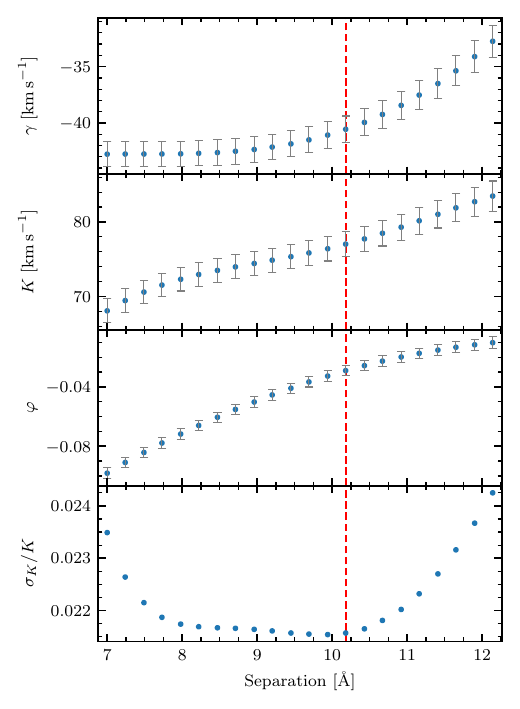}

        \caption{Diagnostic diagram for the H$\alpha$ emission line based on observation from 2019. The red dashed line marks the separation $a = 11.16\, \mathrm{\AA}$ which was used for the radial velocity determination.}
        \label{F:DD}
\end{figure}

For this analysis, the spectra were corrected for the absorption features (see \ref{S:AF}) by re-normalising the spectra to the average absorption profile of the corresponding line.  We did not apply this correction to the spectra of the H$\alpha$ line obtained in 2017, as absorption in these spectra couldn't be reliably estimated. The width of the Gaussian profiles was set to $\sigma = 1.2 \: \mathrm{\AA}$ for spectra of the H$\beta$ line obtained in 2017 and $\sigma = 2 \: \mathrm{\AA}$ for all other instances. The best value for the separation of the two Gaussians was determined from the diagnostic diagrams as the separation, for which the relative error of the semi-amplitude $\frac{\sigma_K}{K}$ has the smallest value. 
In case of H$\beta$ the separation was $a = 10.0 \: \mathrm{\AA}$ for the 2017 spectra and $a = 10.67 \: \mathrm{\AA}$ for the 2019 data, in case of H$\alpha$ the separation was $a = 7.29 \: \mathrm{\AA}$ for the 2017 spectra and $a = 10.18 \: \mathrm{\AA}$ for the 2019 data. The diagnostic diagram for H$\alpha$ line based on the 2019 data is shown in Fig.~\ref{F:DD}.
The separation of the two Gaussians should correspond to the same shift in radial velocity for each line, which is, however, not the case for H$\alpha$ and H$\beta$. The ideal separation for H$\beta$ is larger than for H$\alpha$, contrary to the expectation. This is likely caused by the presence of the absorption component
The presence of the absorption component, which could not be corrected in the case of the 2017 data, results in a smaller separation $a$ compared to that of H$\beta$ Even though the 2019 data of H$\alpha$ were corrected for absorption, its profile could be determined inaccurately due to its low prominence.

\begin{figure}[!tb]
        \centering
        
       \includegraphics[width=0.49\textwidth]{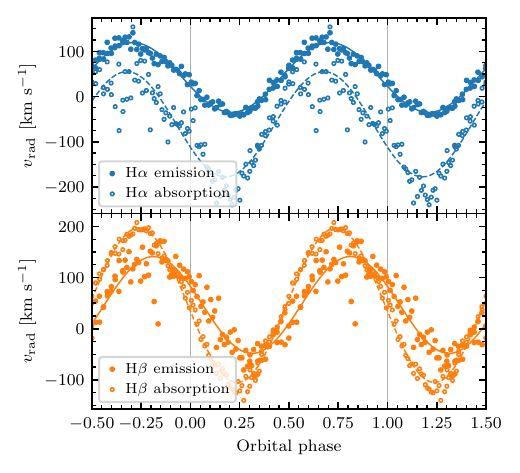}

        \caption{Radial velocity curves based on the broad emission component of H$\alpha$ and H$\beta$ emission lines observed in 2019. Radial velocities derived from the absorption component are plotted for comparison.}
        \label{F:RV_EM01}
\end{figure}

Fig.~\ref{F:RV_EM01} shows the radial velocities based on the wings of H$\alpha$ and H$\beta$ observed in 2019 and compares them with the radial velocities determined from the absorption component. The parameters of the best fits of the derived radial velocity curves are listed in Table~\ref{T:RV_SOLO}.
There is a clear phase offset between the radial velocity curves obtained from the broad emission lines and the ones obtained from the absorption lines. 

This is a known problem of determining radial velocity from the broad emission lines \citep[see][chap.~2.7.6]{1995cvs..book.....W} which is caused by asymmetric brightness distribution. While these usually originate in the accretion disc, bright emission line regions outside the disc - as discussed before - would have the same effect. While the Double-Gaussian method combined with the diagnostic diagram takes care of the typical asymmetric effects coming from the bright spot at the outer rim of the accretion disc, any asymmetry at higher velocities - whether originating from inside the disc or from material further outside bound to the binary orbit will cause a phase shift in the radial velocities derived from the wings of the emission lines. One should also keep in mind that while emission from the inner part of the disc
could contribute to the wings of the broad emission line, the absorption component is observed even further outside at higher radial velocities and therefore offers a more robust way to analyse the
inner part of the disc.
Because of these discussed limitations of the Diagnostic Diagram method, we decided to use the radial velocity curves derived from the strong absorption components in the subsequent analysis.

\subsection{Narrow emission line}

\begin{figure}[tb]
        \centering
        
       \includegraphics[width=0.49\textwidth]{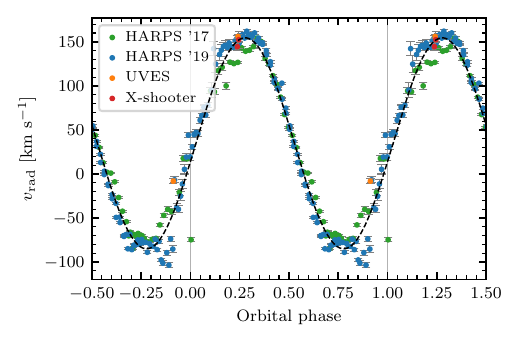}
       
        \caption{Radial velocity of the narrow component of H$\alpha$ based on HARPS, UVES and X-shooter data. The dashed line represents the best fit to the presented data.}
        \label{F:RV_SC}
\end{figure}

Radial velocities of the narrow emission component of the H$\alpha$ line were determined by fitting the spectra with a multi-component model consisting of four Gaussians - one to fit the absorption component, one to fit the narrow emission and two to fit the remaining emission originating in the disc. The radial velocities were then computed from the position of the Gaussian representing the narrow emission. The resulting radial velocity curve is plotted in Fig.~\ref{F:RV_SC}. We observed no significant difference between the radial velocity curves obtained from 2017 and 2019 observations, therefore we fitted all HARPS data simultaneously with a sine function. The parameters of the best fit are given in Table~\ref{T:RV_SOLO}.

The semi-amplitude determined by the fitting is in good agreement with the value presented by \cite{1990A&A...230..326B} and our radial velocity curve is anti-phased to the ones derived from absorption lines and broad emission lines as also found by those authors, which concluded the narrow emission originates from the irradiated side of the secondary star.

The radial velocity curve of the narrow component is in agreement with velocity values derived from the narrow component using the X-shooter and UVES spectra, as can be seen in Fig.~\ref{F:RV_SC}.

\subsection{O-C diagram}

\begin{figure}[tbh]
        \centering

       \includegraphics[width=0.49\textwidth]{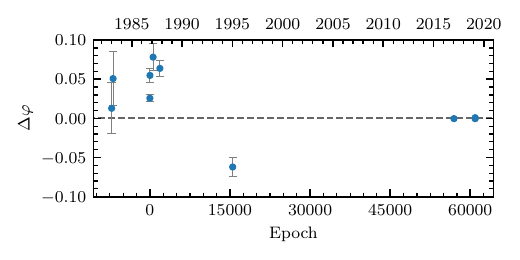}

        \caption{$O-C$ diagram for the times of the zeroth phase derived from the radial velocity curves.}
        \label{F:OC_01}
\end{figure}

We used all Balmer absorption lines except for H$\alpha$ to determine the moments of the zeroth phase for each night. This was achieved by simultaneously fitting the radial velocity curve with the zeroth phase being a shared free parameter for all curves. The derived moments of the zeroth phase are listed in Table~\ref{T:OC_01} together with previously published data. 
As the error of the orbital period determined by \cite{1999AcA....49...73K} is $4 \cdot 10^{-7}\;\mathrm{d}$ and the time elapsed between observations of \cite{1999AcA....49...73K} and HARPS observations corresponds to $\sim 50\,000$ epochs, the zeroth phase for the HARPS  observations could be predicted with an accuracy of $\sim 0.1$ in the orbital phase. This accuracy was high enough to reliably determine the epoch numbers of the HARPS observations. 
We used the data presented in Table~\ref{T:OC_01} to derive an improved spectroscopic ephemeris of the system 
\begin{equation}
\label{E:OCE}
T_0 = \mathrm{HJD}\; 2\,446\,722.8293(17) + 0.19392793(3) \cdot E.
\end{equation}
The $O - C$ diagram of the moments of phase zero constructed with the improved spectroscopic ephemeris is shown in Fig.~\ref{F:OC_01}.

The times of zeroth phases presented in Table~\ref{T:OC_01} were determined from radial velocities obtained with different methods. Radial velocities published by \cite{1983MNRAS.204P..35W} and \cite{1984AJ.....89..389E} were calculated from the central positions of the Balmer hydrogen lines. \cite{1990A&A...230..326B} used a two-component model consisting of two Gaussians to model the emission in H$\alpha$ and  H$\beta$, where one component (sharp) represented the emission originating from the secondary, and one component (broad) represented the emission originating in the disc. \cite{1990A&A...230..326B} then used the radial velocities of the broad component to determine the times of the zeroth phase. \cite{1999AcA....49...73K} used emission lines of hydrogen and \ion{Ca}{II} observed in the spectral range between $8090\;\mathrm{\AA}$ and $9200\;\mathrm{\AA}$ to determine the velocities by fitting each line with a Gaussian. 

The different approaches used to determine the zeroth phase might cause scatter in the $O - C$ diagram which would not reflect real physical changes caused by the dynamics of the system, similarly as we see a phase offset between the radial velocity curves of absorption and emission discussed in Section~\ref{S:BEL}.

\begin{table}[ht]
    \centering
        \caption{Moments of spectroscopic phase zero}
    \label{T:OC_01}
    \centering
    \begin{tabular}{r l l r l}
    \hline\hline  \noalign{\smallskip}
        Cycle       & HJD           &      Error                 &       $O-C$  & Ref.\\
                    & 2400000+      &       [days]               &       [days] &      \\
       \hline  \noalign{\smallskip} 
         $    -7158 $ &     $   45\,334.69570 $ &        $      0.00632 $ &   $  0.00242 $     & (1)\tablefootmark{a} \\
         $    -6895 $ &     $   45\,385.70610 $ &        $      0.00675 $ &   $  0.00978 $     & (2)\tablefootmark{a} \\
         $        0 $ &     $   46\,722.83430 $ &        $      0.00091 $ &   $  0.00493 $     & (3)  \\
         $       20 $ &     $   46\,726.71850 $ &        $      0.00170 $ &   $  0.01058 $     & (3)  \\
         $      603 $ &     $   46\,839.78300 $ &        $      0.00328 $ &   $  0.01509 $     & (3)  \\
         $     1870 $ &     $   47\,085.48690 $ &        $      0.00201 $ &   $  0.01231 $     & (3)  \\
         $    15\,510 $ &     $   49\,730.63940 $ &        $      0.00235 $\tablefootmark{b} &   $ -0.01209 $     & (4)  \\
         $    56\,943 $ &     $   57\,765.66715 $ &        $      0.00002 $ &   $ -0.00007 $     & (5)  \\
         $    60\,914 $ &     $   58\,535.75493 $ &        $      0.00002 $ &   $ -0.00008 $     & (5)  \\
         $    60\,919 $ &     $   58\,536.72481 $ &        $      0.00002 $ &   $  0.00016 $     & (5)  \\
        
    \hline
    \end{tabular}
    
    \tablefoot{Errors refer to $1\sigma$. References: (1) - \cite{1983MNRAS.204P..35W}, (2) - \cite{1984AJ.....89..389E}, (3) - \cite{1990A&A...230..326B}, (4) - \cite{1999AcA....49...73K}, (5) - this work\\
    \tablefoottext{a}{Times of the zeroth phase were determined by \cite{1990A&A...230..326B} using data presented in referenced works. }\\
    \tablefoottext{b}{The error was not given by the authors, the value listed in the table was determined by least-square fitting of the radial velocities published by authors.}
    }

\end{table}{}

\section{Doppler tomography}

\begin{figure*}[tbh]
        \centering

       \includegraphics[width=0.49\textwidth]{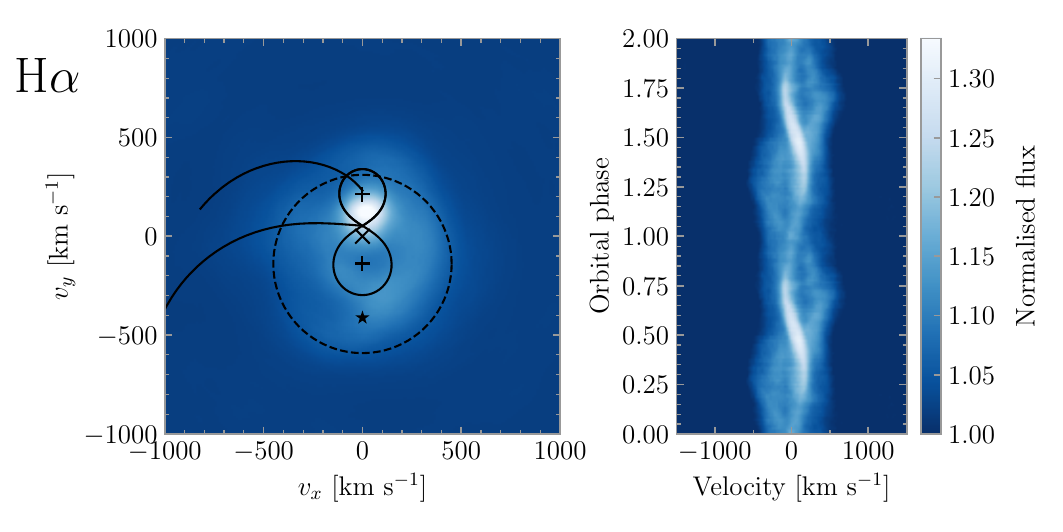}
       \includegraphics[width=0.49\textwidth]{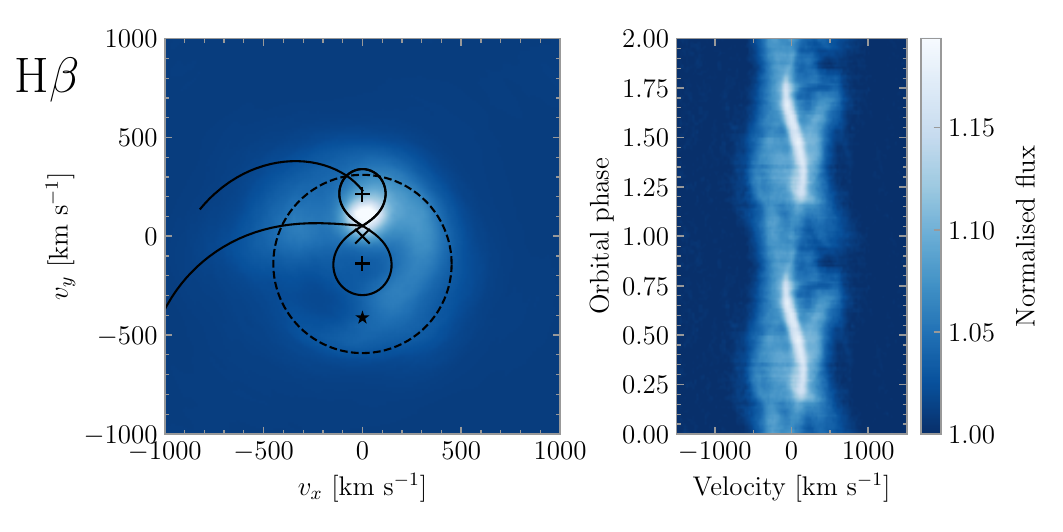}
       
       \includegraphics[width=0.49\textwidth]{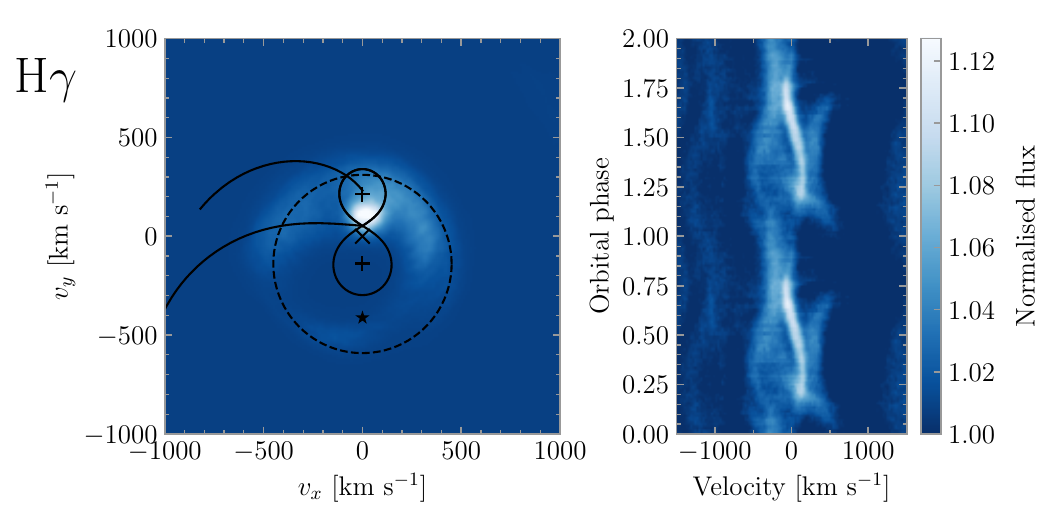}
       \includegraphics[width=0.49\textwidth]{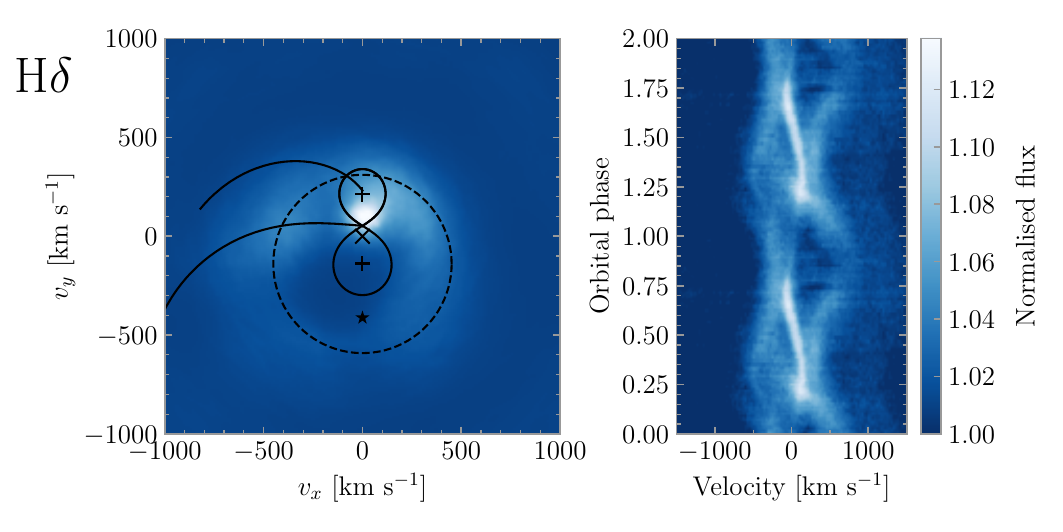}
       
       \includegraphics[width=0.49\textwidth]{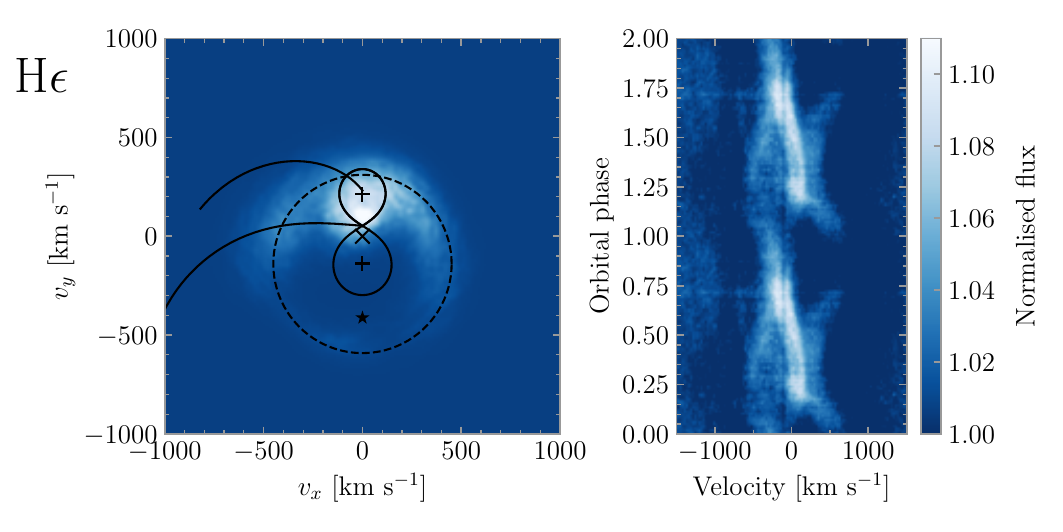}
       \includegraphics[width=0.49\textwidth]{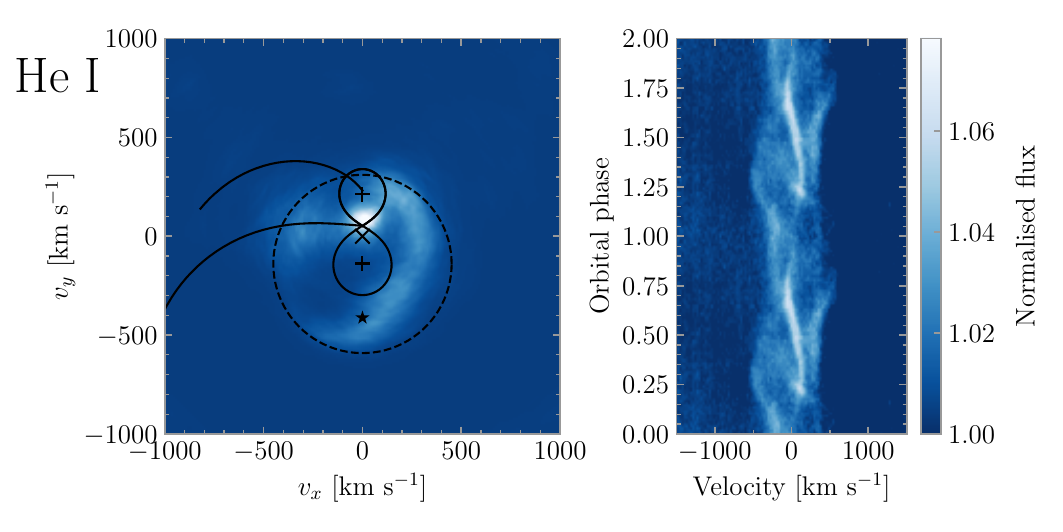}
       
       \includegraphics[width=0.49\textwidth]{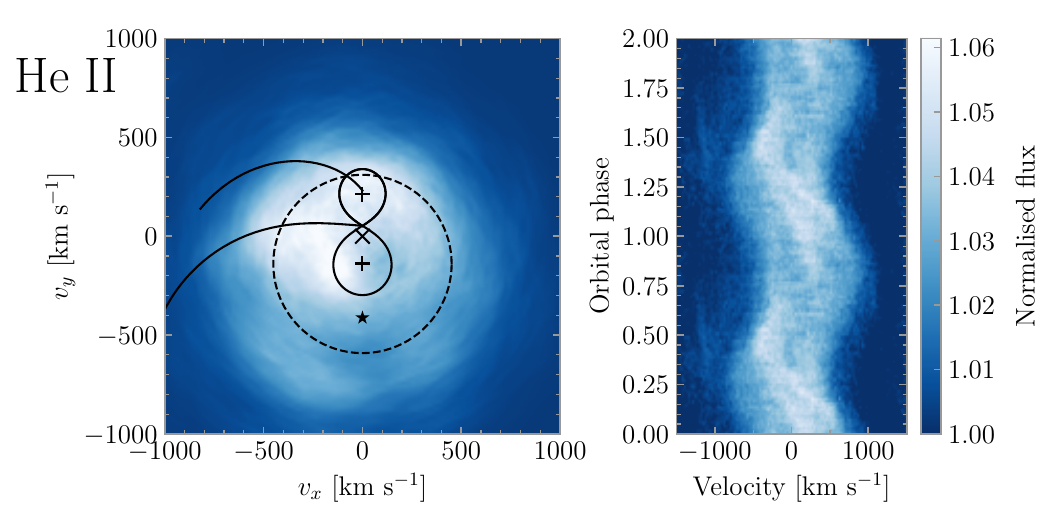}
       \includegraphics[width=0.49\textwidth]{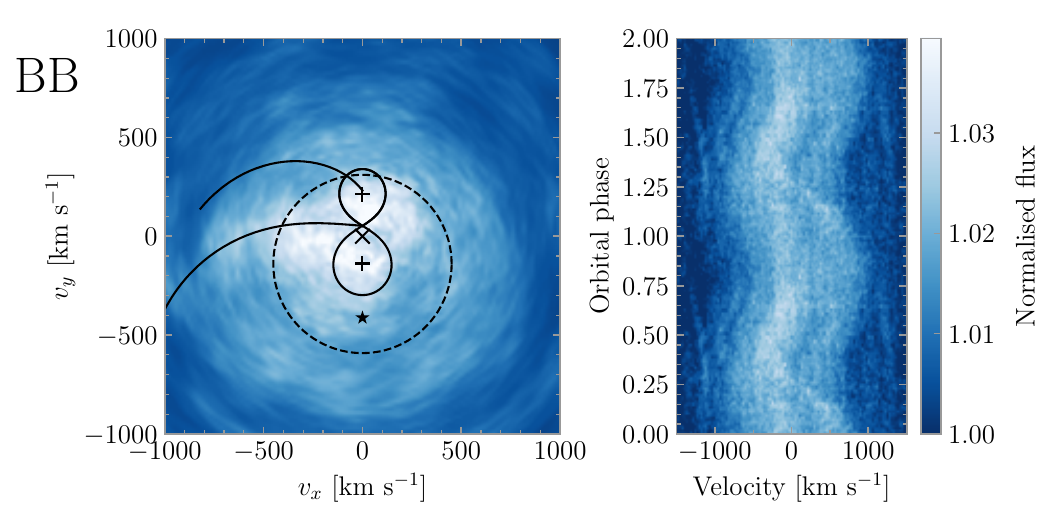}

        \caption{Doppler tomography based on HARPS observations from all three nights, the colour of the Doppler maps corresponds to arbitrary units of emission intensity. The positions of the primary and secondary are marked by plus signs, the centre of mass is marked by a cross sign, and the position of the $\mathrm{L}_3$ point is marked by the star symbol. The velocity of the $\mathrm{L}_3$ point corresponds to its velocity caused by the rotation of the binary system; the keplerian velocity of the disc extrapolated to this location is larger by about $40\;\mathrm{km}\,\mathrm{s}^{-1}$.       
        Roche lobes of both components and the expected stream trajectory are outlined by solid lines, the tidal limitation radius is outlined by the dashed circle. Doppler maps were computed from lines H$\alpha$, H$\beta$, H$\gamma$, H$\delta$, H$\varepsilon$, \ion{He}{i} $\lambda6677$ \AA, \ion{He}{ii} $\lambda4686$ \AA\ and from Bowen blend (BB).  All maps derived from the analysis of the Balmer lines are corrected for the presence of the absorption features, as described in Section~\ref{S:AF}. For each map a trailed spectrum of the corresponding line is shown. Note that the scale of the Doppler maps was chosen to highlight the observed structures and does not correspond to the scale of trailed spectra.
        Fig.~\ref{F:DT_LOG} shows the same Doppler maps, only in inverse hyperbolic sine scale, in which the low-amplitude features are higlihted.}
        \label{F:DT_01}
\end{figure*}

We used Doppler tomography to map the accretion disc in velocity coordinates $(v_x,v_y)$. This technique was developed by \cite{1988MNRAS.235..269M} and uses time-resolved spectroscopic observations to compute maps of the system, which display the intensity of the emission line emitted at velocity coordinate $(v_x,v_y)$. We employed the code by \cite{1998astro.ph..6141S} which applies the maximal entropy method to compute the Doppler maps. We used {\sc Python3} programming language for plotting the maps. The Roche lobe, stream, and position of the components overlaid over the maps were plotted using the {\sc Python3} package {\sc pydoppler}\footnote{Available at \url{github.com/Alymantara/pydoppler}} by \cite{2021ascl.soft06003H}. When plotting the Roche lobe and the accretion stream, we assumed the mass of the primary $M_{\mathrm{WD}} = 0.8 \:\mathrm{ M}_{\odot}$ and the mass of the secondary $M_{2} = 0.52 \:\mathrm{ M}_{\odot}$, which were derived by \cite{2007ApJ...662.1204L}. This gives the mass ratio $q = 0.65$.

We binned the spectra used for the computation by a factor of 30 which filtered out the noise present in the spectra but still provided a spectral resolution high enough to study the structure of the system. We combined all HARPS spectra when computing the maps, as maps computed for individual nights did not show any significant changes of the structure, which was also the case for the individual spectra as can be seen in Fig.~\ref{F:03}. The orbital phases of the spectra were computed using the Equation~\ref{E:OCE}, which proved to be precise enough to combine data obtained two years apart. We computed maps for emission lines with sufficient signal-to-noise ratio,
all Doppler maps are presented in Fig.~\ref{F:DT_01} and the same maps but in inverse hyperbolic sine scale are shown in Fig.~\ref{F:DT_LOG}. The spectra of H$\alpha$ line were contaminated by telluric lines and the spectra of H$\varepsilon$ were contaminated by interstellar lines, which would create fake circular structures in the Doppler maps of these lines. We have therefore corrected the spectra for the contamination by normalising them to an average spectrum of contaminating lines and the Doppler maps presented are free of these fake structures.

The Doppler maps of the hydrogen and \ion{He}{i} $\lambda6677$~\AA\ lines show a bright emission corresponding to the velocity coordinates of the part of the secondary adjacent to the Lagrangian L$_1$ point. This emission is also the dominant feature in the trailed spectra of these lines with maximal brightness around phase $\varphi = 0.5$, which corresponds to the face-on view of the irradiated part of the secondary. In spectra obtained close to phase $\varphi = 0.0$, when the irradiated part of the secondary is hidden out-of-view, we see only little to no emission  originating from that region, however, in some cases (e.g. the H$\alpha$ line) it doesn't disappear completely. 

Doppler maps of H$\alpha$, H$\beta$ and \ion{He}{i} $\lambda6677$~\AA\  show an arm-like feature in the lower-right quadrant ($v_x = 200 \; \mathrm{km \; s}^{-1}, v_y = -250 \; \mathrm{km \; s}^{-1}$), which is located outside of the tidal-limitation radius. This feature corresponds to the position of an outflow from the accretion disc in the work by \cite{2010MmSAI..81..187B}. This outflow gives rise to a gaseous circum-binary envelope and the model by \cite{2010MmSAI..81..187B} shows the presence in a region outside the disc of matter  connecting to a bow shock, which forms as the binary moves through the circum-binary envelope. The emission from this outflow region as computed by \cite{2010MmSAI..81..187B} is located in our Doppler maps at coordinates of $v_x \gtrsim 0 \; \mathrm{km \; s}^{-1}$ outside of the accretion disc. Similar structures have also been observed in  other NLs, e.g. RW~Sex, 1RXS J064434.5+334451  \citep{2017MNRAS.470.1960H} and RW~Tri \citep{2020MNRAS.497.1475S}.

The Doppler maps of the Balmer lines show an emission region coinciding with the position of the tidal-limitation radius, even though in the case of H$\alpha$ and H$\beta$ it is less prominent when compared with other features in those maps. We interpret this feature as the emission from the outer rim of the accretion disc. It does not have the usual doughnut shape which is typical for accretion discs of CVs, as the emission intensity is not circularly symmetric with the lower part exhibiting weaker emission. This could be caused by a matter in the outflow region blocking the view of the accretion disc.

Doppler maps of \ion{He}{ii} $\lambda 4686$ and the Bowen blend show the highest intensity at low velocities outside of the accretion disc - the map of \ion{He}{ii} has highest intensity at coordinates ($v_x \sim -200 \; \mathrm{km \; s}^{-1}, v_y \sim -50 \; \mathrm{km \; s}^{-1}$) and the maximal intensity in the map of Bowen blend is centred on the centre of mass of the system. The low velocities suggest that these lines are formed in wind from the accretion disc. This is similar to the NL SW Sex, which also shows low-velocity emission in \ion{He}{ii} $\lambda 4686$ and the Bowen blend which \cite{1986ApJ...302..388H} interpreted as wind emission. Similarly, observation of DN IP Peg in outburst presented by \cite{1990ApJ...349..593M} show a low-velocity \ion{He}{ii} $\lambda 4686$ emission and its Doppler maps has maximal intensity at a similar position as the one of IX Vel.

\section{Doppler map of near-infrared emission lines}

\begin{figure}[!tb]
        \centering
        
        \includegraphics[width=0.49\textwidth]{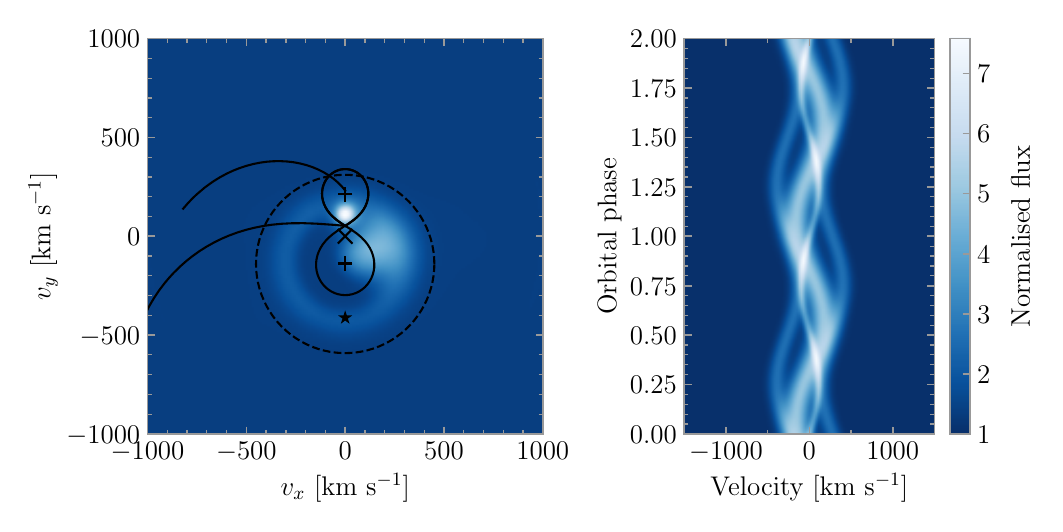}

        \caption{Doppler map (\textit{left}) and trailed spectra (\textit{right}) based on the three-component model of emission lines presented by \cite{1999AcA....49...73K}.}
        \label{F:DT:K}
\end{figure}

\cite{1999AcA....49...73K} analysed time-resolved near-infrared spectra and fitted the observed spectral lines with a three-component model.
Even though they did not present a Doppler map based on their time-resolved spectra, their three-component model allowed us to reconstruct a Doppler map of IX~Vel. We used the parameters presented by \cite{1999AcA....49...73K} in their table 4 and relative strengths of \ion{Ca}{ii} $\lambda 8542$ given in their table 3 to compute a set of phase-resolved spectra. We subtracted a value of $0.13$ from the phase parameter of each component so that the parameter $\Phi_2 = 0$ and the zeroth phase is the moment of the inferior conjunction of the secondary.
We then used the spectra to compute a Doppler map, the resulting map is presented in Fig.\ref{F:DT:K}. 

The map shows that the components, which \cite{1999AcA....49...73K} call disc component and  hot spot component, lie outside of the tidal limitation radius of the accretion disc, and that the hot spot component is not located in the usual position of the bright spot. Therefore, the two components are more likely linked to an outflow material from the system.
All the emission from the model originates at velocities lower than $500 \; \mathrm{km \; s}^{-1}$ which resembles several maps presented in Fig.~\ref{F:DT_01}, especially the map of H$\alpha$ for which most of the emission can be found at lower velocities. The hot spot component of Kubiak's model is located at coordinates ($v_x = 200 \; \mathrm{km \; s}^{-1}, v_y = -50 \; \mathrm{km \; s}^{-1}$) and coincides with an emission source found in our Doppler maps, most noticeably in the map of H$\gamma$.

\section{Temperature distribution map}

\begin{figure*}[!tbh]
        \centering

       \includegraphics{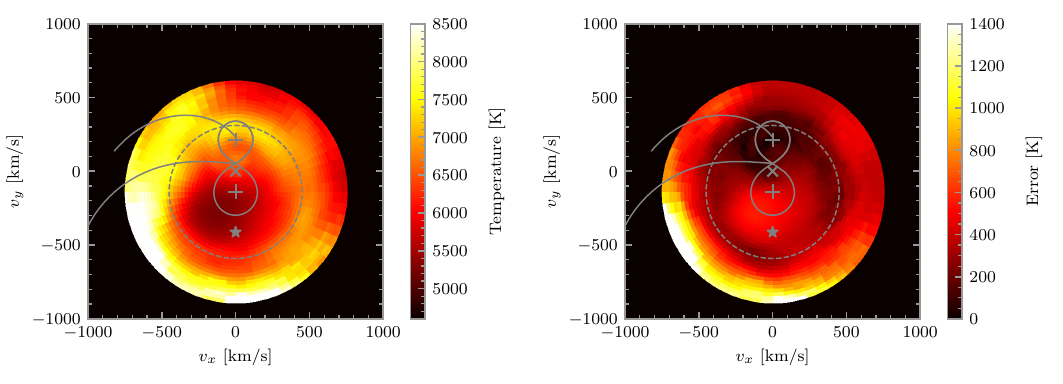}

        \caption{Temperature map based on Doppler maps of lines H$\alpha$, H$\beta$, H$\delta$, H$\varepsilon$ (\textit{left panel}) and the estimated error of the temperature map (\textit{right panel}).
        }
        \label{F:DT_T}
\end{figure*}

We have obtained Doppler maps of five Balmer lines, which show the intensity of the corresponding line in velocity coordinates. This intensity is dependent on the hydrogen density in the disc as well as on the temperature of hydrogen, but a Doppler map based on a single line is not sufficient to distinguish between temperature-related structures and density-related structures. We used the Doppler maps to compute an approximate temperature distribution map following a method which was used by \cite{2016MNRAS.463.3290R} to compute a temperature distribution map of the accretion disc of the dwarf nova V2051~Oph. This method uses the assumption that the spectrum emitted by the accretion disc can be modelled by an isothermal and isobaric pure-hydrogen slab whose flux can be described by a function \begin{equation}
    I_{\lambda}\left( T \right) = B_{\lambda}\left( T \right)\left( 1 - \mathrm{e}^{- \tau_{\lambda}} \right),
\end{equation} where $B_{\lambda}\left( T \right)$ is the Planck function describing the blackbody radiation and $\tau_{\lambda}$ is the wavelength dependant optical depth. The method further assumes that the optical depth in emission-lines is centred at $\tau_{\lambda} > 1$ and is significantly higher than in the continuum. Under this assumption, the decrements of the Balmer lines should be described by a Planck function and the temperature of an element with coordinates $(v_x,v_y)$ can be estimated by fitting the flux values of corresponding elements of different Doppler maps with a black-body. 

We used the Doppler maps of the H$\alpha$, H$\beta$, H$\delta$, and H$\varepsilon$ emission lines to construct a map of the temperature distribution for the disc in IX~Vel. We have excluded the map of the H$\gamma$ emission line, because the absorption component of this line was too complex to be sufficiently fitted by a single Gaussian, probably due to blending with metal lines in its vicinity. As a result, the Doppler map based on the reconstructed emission component could represent incorrect flux values unsuitable for the estimation of the temperature distribution. 

The Doppler maps presented in Fig.~\ref{F:DT_01} were produced from normalised spectra, therefore for the temperature map computation we scaled each map by the flux of the continuum at the position of each corresponding line. The scaling factors were derived by fitting the flux-calibrated X-shooter spectra of IX~Vel with a Planck function. 

We decided to compute the temperature distribution map for a grid with pixels of a larger size than the ones of the Doppler maps in order to minimise the effect of noise in the Doppler maps affecting the temperature distribution values. Averaging over a larger area is important, especially for such $(v_x,v_y)$ where the intensity of the emission lines is low. As can be seen on the Doppler maps in Fig.~\ref{F:DT_01}, low-intensity regions are located predominantly at high velocities. We have therefore decided to use a polar grid for the temperature distribution map with its centre located at the position of the WD. With this setup, the temperature for higher velocities is based on more pixels of the Cartesian grid and the symmetry of the polar grid reflects the shape of a circular accretion disc with its centre at the WD position. We also excluded pixels with high-velocity coordinates, which did not contain enough signal for a reliable fit, and we computed the temperature distribution map only for a circular mask with a radius of $v = 750\:\mathrm{km}\:\mathrm{s}^{-1}$ centred on the WD. The mask was divided into 60 annuli and 30 sectors.

The resulting temperature distribution map is shown in Fig.~\ref{F:DT_T} together with a map showing the corresponding errors. 
The map itself should be considered only a rough approximation of the actual temperature distribution, as numerous assumptions about the origin of the emission lines were used. For this reason, we also do not interpret the temperature values as absolute ones, but rather as relative values used to distinguish various features in the system.

The area of the map with the highest temperature (($T > 8500 \; \mathrm{K}$) is located at the high velocities in the lower-left quadrant ($v_x = -600 \; \mathrm{km \; s}^{-1}, v_y = -400 \; \mathrm{km \; s}^{-1}$) and extends to the bottom part of the map. This area, however, also has the largest error estimates and therefore its existence is dubious. Another region of high temperature ($T \simeq 7500 \; \mathrm{K}$) is in the upper-left quadrant corresponding to the expected position of a bright spot. Similarly to the Doppler maps, structures corresponding to the tidal-limitation radius are present, but are not circularly symmetrical, with the rim of the disc in the lower-left quadrant being colder than the rest of the disc. 
The temperature along the tidal limitation radius varies between $T \simeq 6300\,\mathrm{K} $ and $T \simeq 7200\,\mathrm{K}$ with an average temperature of about $T \simeq 6900\,\mathrm{K}$.
The irradiated face of the secondary is also present in the temperature distribution map ($T \simeq 6700\,\mathrm{K}$), although it is a less prominent feature than in the Doppler maps presented in Fig.~\ref{F:DT_01}.
Even if the absolute values are not to be trusted, we can say that in relation to each other, the bright spot area is about $600\; \mathrm{K}$   hotter than its surroundings. This is consistent with the findings of \cite{2007ApJ...662.1204L}, that the bright spot can be only about $500\;\mathrm{K}$ hotter than the rest of the rim of the disc.

\section{IX Vel as an RW Sex star}

The Doppler maps of IX Vel resemble the maps of RW Sex and 1RXS~J064434.5+334451 constructed by \cite{2017MNRAS.470.1960H}. Both systems are NL stars and exhibit a strong emission from the illuminated face of the secondary and an additional emission corresponding to the position of an outflow zone.
Similar features are also observed in RW Tri \citep{2020MNRAS.497.1475S} and BG Tri \citep{2021MNRAS.503.1431H}, 
which prompted Hernandez to group them together with RW Sex as the prototype.
IX Vel fits well within these systems and can just be considered belonging to the group of RW Sex stars as defined by \cite{2017MNRAS.470.1960H}.

However, many other NLs show one or the other of these features:
\cite{2004MNRAS.353.1135T} presented Doppler tomography of AC Cnc and V363 Aur, both of which show emission from the secondary close to the $\mathrm{L}_1$ point and low-velocity emission which appears to originate outside of the accretion disc. In the case of AC Cnc, the low-velocity emission is circularly symmetrical and could also be related to a wind originating in the disc. For V363 Aur, the low-velocity emission is located at negative $v_y$ velocities and strongly resembles emission from an outflow zone described in the previously mentioned systems. Low-velocity emission which seems to originate outside of the accretion disc can be also found in Doppler maps of Balmer lines of SW Sex and DW UMa constructed by \cite{2013MNRAS.428.3559D}, even though in the case of SW Sex, the emission lies close to the tidal limitation radius and could therefore originate in the disc as well. These two systems do not show emission originating at the secondary, but they show emission linked to the bright spot and in the case of SW Sex also emission linked to the gas stream. \cite{2013MNRAS.428.3559D} also constructed Doppler maps for \ion{He}{ii} for both SW Sex and DW Uma. While their \ion{He}{ii}-map of SW Sex looks similar to the maps of the Balmer lines, the map of DW UMa shows emission centred on the WD which likely originates in an accretion disc wind. The latter was also recently reported in system ASAS~J071404+7004.3 by \cite{2022MNRAS.510.3605I}.
Does this mean that all these stars are potential members of the RW-Sex group? It seems that many of these systems also belong to the SW Sex stars or are at least NLs with high mass transfer rates. The presence of an outflow as well as the strong irradiation of the secondary might be expected for such systems and would thus be another indication for a high mass transfer system. 
SW Sex stars show a strong emission originating in the bright spot and an absorption feature near the orbital phase $\varphi = 0.5$, which \cite{2014AJ....147...68T} interprets as evidence of the outflow zone. 
In the case of long-period NLs ($P_{\mathrm{orb}} > 4\,\mathrm{h}$), the outflow region is observed as a strong emission source and the bright spot is not a prominent feature in these systems \citep{2017MNRAS.470.1960H}.

The presence of outflows in NLs would play an important role in understanding the evolution of CVs, as it would provide an additional mechanism for orbital angular momentum loss, acting alongside the traditionally considered magnetic braking and gravitational wave radiation. \cite{2013CEAB...37..361S} finds that almost all CVs with orbital periods in a range between $2.8\;\mathrm{h}$ and $4\;\mathrm{h}$ are NLs of SW Sex type and concludes that they must represent an evolutionary phase of CVs as all long-period CVs will eventually travel through this period range. The few systems in this period range for which the secular mass transfer was measured \citep[see][and references therein]{2017MNRAS.466.2855P} show high accretion rates, much higher than predicted from the models \citep{2011ApJS..194...28K}. This indicates that indeed additional braking mechanisms are needed to explain these systems, and several have been dicussed in the literature \citep{1997MNRAS.289...59Z, 2009ApJ...693.1007T, 2011ApJS..194...28K}. The presence of outflows as observed in IX Vel and the afore mentioned systems (see Table~\ref{tab:CV:COMP}) could just be another option.

\begin{table}[]
    \caption{Examples of single-peaked NLs with similar characteristics to IX Vel}
    \centering
    \begin{tabular}{l l c c   r }
    \hline \hline
        Name                    & $P_\mathrm{orb}$ & $i$                  & Reference\\
    \hline 
        IX Vel         & $4.6\;\mathrm{h}$& $60^{\circ}$   & (1)                  \\
        ASAS J071404+7004.3     & $3.3\;\mathrm{h}$& $60^{\circ}$   & (2) \\
        RW Tri                  & $5.6\;\mathrm{h}$& $77^{\circ}$       & (3) \\
        BG Tri                  & $3.8\;\mathrm{h}$& $25^{\circ}$   & (4) \\
        RW Sex                  & $5.9\;\mathrm{h}$& $34^{\circ}$   & (5) \\
        1RXS J064434.5+334451   & $6.5\;\mathrm{h}$& $74^{\circ}$   & (5) \\
        AC Cnc                  & $7.2\;\mathrm{h}$& $76^{\circ}$   &  (6)\\
        V363 Aur                & $7.7\;\mathrm{h}$& $70^{\circ}$   & (6) \\
        SW Sex                  & $3.2\;\mathrm{h}$& $79^{\circ}$   & (7, 8)  \\
        DW UMa                  & $3.3\;\mathrm{h}$& $82^{\circ}$   & (8) \\
    \hline
    \end{tabular}
    \tablefoot{ References: (1) this work, (2) \cite{2022MNRAS.510.3605I},  (3) \cite{2020MNRAS.497.1475S}, (4) \cite{2021MNRAS.503.1431H}, (5) \cite{2017MNRAS.470.1960H}, (6) \cite{2004MNRAS.353.1135T}, (7) \cite{1997MNRAS.291..694D} (8) \cite{2013MNRAS.428.3559D}.}
    \label{tab:CV:COMP}
\end{table}

\section{Conclusion}

We analysed  the time-resolved spectroscopy of IX~Vel which was obtained  during two different runs, one in 2017 and one in 2019. While the general structure of spectra was the same during both runs, a comparison with spectra obtained on other occasions showed that it is not a general rule for this system. Namely, the hydrogen emission line profiles observed in 2017 and 2019 significantly vary from the ones obtained few years before, in 2014. 

We used time-resolved spectroscopy to determine the radial velocity curves of several features of spectral lines and to improve the spectroscopic ephemeris of the system. 

We constructed Doppler maps of the system using the emission lines of hydrogen, helium, and Bowen blend. The maps of hydrogen and \ion{He}{i} show a strong emission originating from the irradiated surface of the secondary, emission from the outer part of the accretion disc and emission originating outside of the disc, which we interpret as a possible outflow from the disc. Maps of \ion{He}{ii} and Bowen blend show emission at low velocities, which we interpret as emission from the accretion disc winds. We conclude that IX Vel can be considered a member of the RW Sex stars as defined by \cite{2017MNRAS.470.1960H}.
Although some maps show a hint of a bright spot emission, most notably the map of \ion{He}{i}, 
there is no strong emission originating from its position.

We used the Doppler maps of H$\alpha$, H$\beta$, H$\delta$, H$\varepsilon$ to compute a temperature distribution map of the system. Due to the strong assumptions used for the computation, one should view the resulting map only as a relative temperature distribution in the system.
The map shows high-temperature features which can be interpreted as the bright spot of the accretion disc, the rim of the accretion disc and the irradiated face of the secondary. In agreement with \cite{2007ApJ...662.1204L} we find temperatures of the bright spot to be about $600\;\mathrm{K}$ brighter than the rest of the accretion disc rim.

The emission originating in the outflow zone, which we see in the Doppler maps of IX Vel, was observed also in other NLs which show signs of high mass transfer.
As the velocities corresponding to the outflow zone are low, its structure in Doppler maps can only be probed using high-resolution time-resolved spectra. 
To further study the plenitude of such outflows, Doppler tomography of the known
high mass-transfer NL systems based on high-resolution spectra are needed to resolve the structure of these low-velocity components in the emission lines and to establish a census of its occurrence.

\begin{acknowledgement}
We are grateful to the anonymous referee for providing us with useful comments and suggestions that improved our manuscript.
We thank Marek Wolf for useful discussions and comments.
Based on observations collected at the European Organisation for Astronomical Research in the Southern Hemisphere under ESO programmes 60.A-9700, 094.D-0344, 69.C-0171 and 088.D-0537.
Based on data obtained from the ESO Science Archive Facility with DOIs: https://doi.org/10.18727/archive/33, https://doi.org/10.18727/archive/50, https://doi.org/10.18727/archive/71. 
This research has made use of ESASky, developed by the ESAC Science Data Centre (ESDC) team and maintained alongside other ESA science mission's archives at ESA's European Space Astronomy Centre (ESAC, Madrid, Spain).
This research is based on observations made with the NASA/ESA Hubble Space Telescope obtained from the Space Telescope Science Institute, which is operated by the Association of Universities for Research in Astronomy, Inc., under NASA contract NAS 5–26555. These observations are associated with program 14637.
This research was supported by the Ministry of Education, Youth and Sports (Czech Republic).
\end{acknowledgement}

\bibliographystyle{aa} 
\bibliography{Yourfile} 

\begin{appendix}

\begin{figure*}[h!]
\section{Doppler maps in inverse hyperbolic sine scale}
        \centering

       \includegraphics[width=0.49\textwidth]{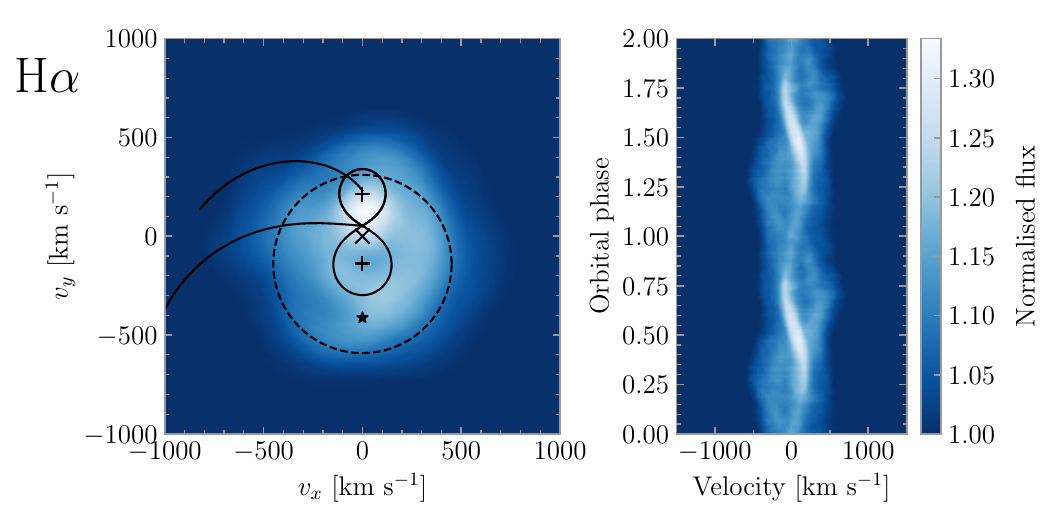}
       \includegraphics[width=0.49\textwidth]{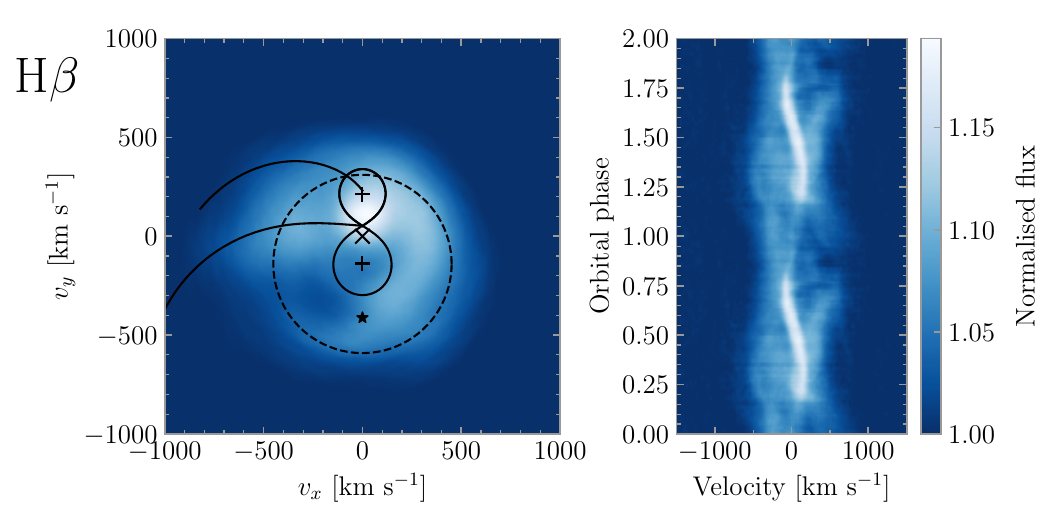}
       
       \includegraphics[width=0.49\textwidth]{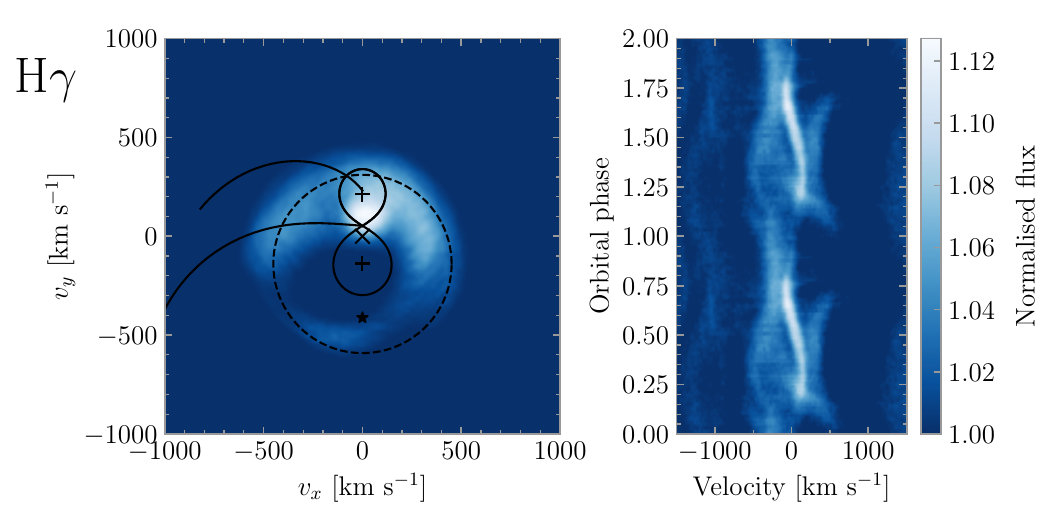}
       \includegraphics[width=0.49\textwidth]{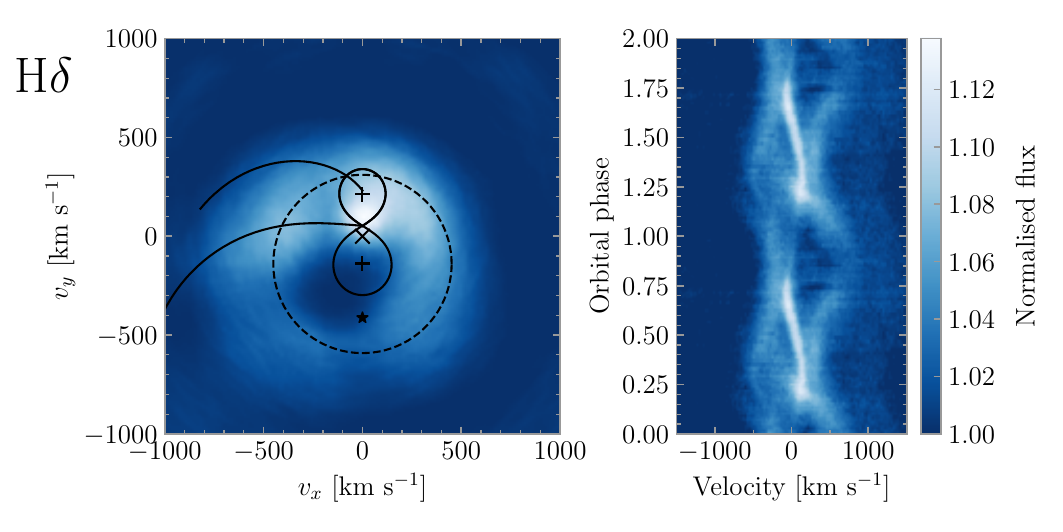}
       
       \includegraphics[width=0.49\textwidth]{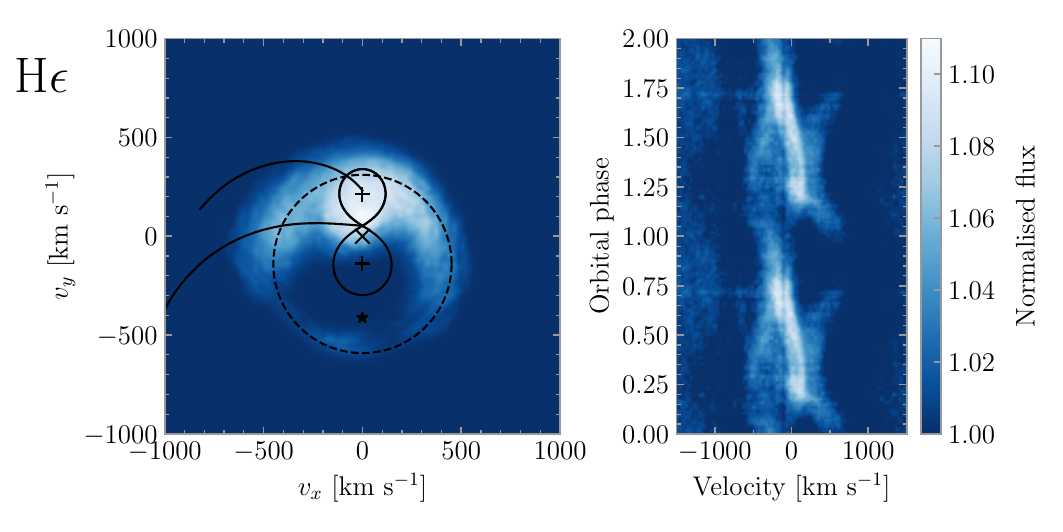}
       \includegraphics[width=0.49\textwidth]{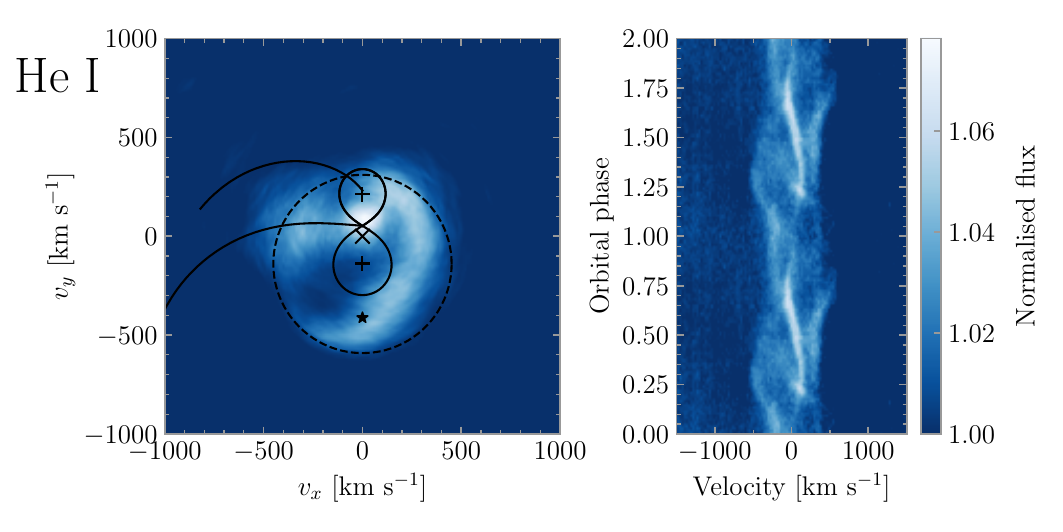}
       
       \includegraphics[width=0.49\textwidth]{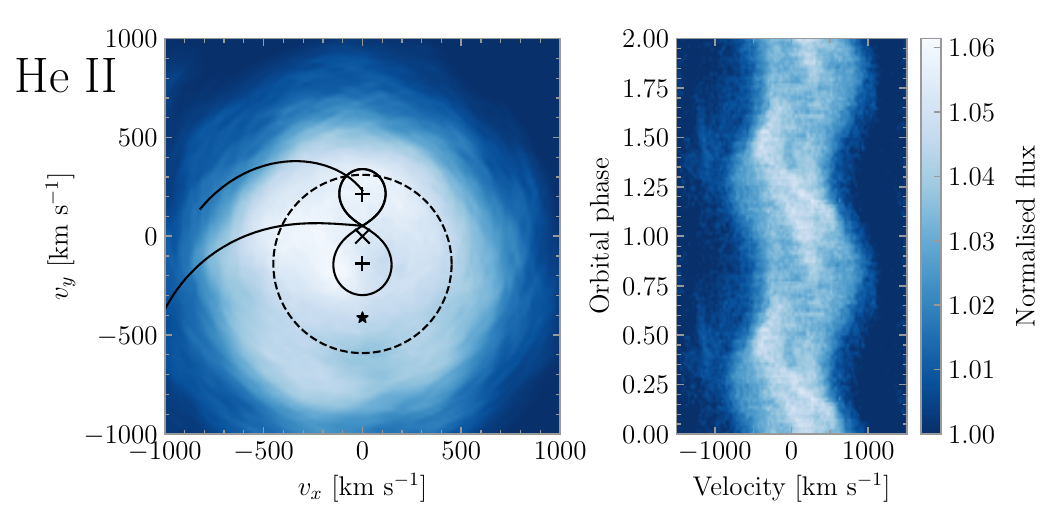}
       \includegraphics[width=0.49\textwidth]{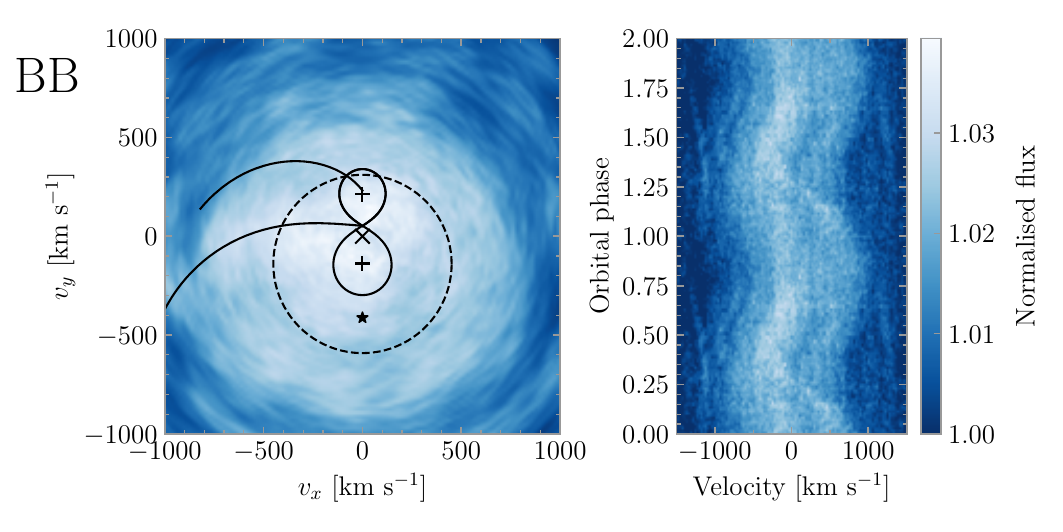}

        \caption{Same as Fig.\ref{F:DT_01}, only the Doppler maps 
        are shown in inverse hyperbolic sine scale to highlight low-amplitude features. 
        }
        \label{F:DT_LOG}
\end{figure*}

\end{appendix}

\end{document}